\documentclass[11pt]{elsarticle}

\usepackage{amsmath}
\usepackage{amssymb}
\usepackage{amstext}
\usepackage{appendix}

\usepackage{graphicx}
\usepackage{multirow}

\usepackage{color} 
\usepackage[numbers]{natbib}
\biboptions{longnamesfirst,semicolon}

\usepackage{url}

\usepackage[caption=false]{subfig}
\usepackage{epsf}
\usepackage{epsfig}
\usepackage{color} 

\usepackage{bm}
\newcommand{\vect}[1]{\boldsymbol{\mathbf{#1}}}

\DeclareMathAlphabet{\mat}{OT1}{cmss}{bx}{n}

\def\longrightharpoonup{\relbar\joinrel\rightharpoonup}
\def\longleftharpoondown{\leftharpoondown\joinrel\relbar}

\def\longrightleftharpoons{
  \mathop{
    \vcenter{
      \hbox{
      \ooalign{
        \raise1pt\hbox{$\longrightharpoonup\joinrel$}\crcr
	  \lower1pt\hbox{$\longleftharpoondown\joinrel$}
	  }
      }
    }
  }
}

\newcommand{\rates}[2]{\displaystyle
  \mathrel{\longrightleftharpoons^{#1\mathstrut}_{#2}}}

\title{Model of haplotype and phenotype in the evolution of a duplicated autoregulatory activator}
\author{Srinandan Dasmahapatra\\Electronics and Computer Science\\Institute for Life Sciences\\Faculty of Physical and Applied Sciences\\University of Southampton, SO17 1BJ, UK\\\emph{email:}{\tt sd@ecs.soton.ac.uk}}
\date{November 28, 2012} 

\journal{Journal of Theoretical Biology}

\begin{document}
\begin{frontmatter}

\begin{abstract}
Gene duplication is believed to play a major role in the evolution of genomic complexity.  The presence of a duplicate frees a gene from the constraint of natural selection, leading to its loss of function or the gain of a novel one.  Alternately, a pleiotropic gene might partition its functions among its duplicates, thus preserving both copies.  Such arguments invoking duplication for novelty or specialisation are not true of diploid genotypes, but only of haplotypes.  In this paper, we study the consequences of regulatory interactions in diploid genotypes  and explore how the context of allelic interactions 
gives rise to dynamical phenotypes that enable duplicate genes to spread in a population.  The regulatory network we study is that of a single autoregulatory activator gene, and the two copies of the gene diverge either as alleles in a diploid species or as duplicates in haploids.
These differences are in their transcriptional ability -- either via alterations to its activating domain, or to its \emph{cis}-regulatory binding repertoire.  When \emph{cis}-regulatory changes are introduced that partition multiple regulatory triggers among the duplicates, it is shown that mutually exclusive expression states of the duplicates that emerge are accompanied by a back-up facility: when a highly expressed gene is deleted, the previously unexpressed duplicate copy compensates for it.  The diploid version of the regulatory network model can account for allele-specific expression variants, and a model of inheritance of the haplotype network enables us to trace the evolutionary consequence of heterozygous phenotypes. This is modelled for the variations in the activating domain of one copy, whereby stable as well as transiently bursting oscillations ensue in single cells.  The evolutionary model shows that these phenotypic states accessible to a diploid, heterozygous genotype enable the spread of a duplicated haplotype.  
\end{abstract}
\end{frontmatter}
\section{Introduction}

Gene duplication is a major source of genomic expansion and is believed to underlie the evolution of complex biological functions \cite{ohnobook70,lynchbook}.  Functions encoded by two copies of the same gene are redundant making the loss of function of a duplicate by mutation a likely outcome \cite{Clark-duplication1994}.  Hence, the observed abundance of duplicates in plant and animal genomes makes the retention of duplicate genes a much studied problem.  The loss of selection pressure on the duplicate can be viewed not as a prerequisite for its elimination, but as an opportunity for it to become more abundant by acquiring a new and fitness-enhancing function \cite{ohnobook70}, a process called neofunctionalization.  If the duplicated gene already had multiple roles (is pleiotropic), their  partitioning among the duplicates would make each essential, a model called subfunctionalization \cite{force-genetics1999}.  Appeals to variations amongst duplicate genes can, in diploid species, also be applied to allelic variations in singleton genes \cite{proulx2006}.   The loss of one or both allelic function has been the basis of debates on whether dominance in Mendelian inheritance is an evolutionary or physiological phenomenon \cite{bourguet99}.  Dominance and gene duplication have both been framed \cite{kondrashov-koonin04} as phenomena that involve gene dosage  --  the contribution of the number of functional genes to phenotype -- and the functional redundancy and fitness of genes may involve quantitative factors.  Quantitative considerations include the disruption of stoichiometrically balanced protein levels when gene duplication increases the expression of one interacting partner\cite{heterosis2010}.  It is the network of interactions that mediate the causal pathways from genes to phenotype and consequent evolutionary outcomes.

Novelties in evolution often emerge via changes to an organism's development.  A common mechanism in developmental trajectories is the transformation of transient stimuli into steady-state expression levels \cite{odom-hepatocyte2006,davidsonbook2006, hobert-terminalselectorPNAS08, desplan-annurev2010} by gene regulatory circuits that implement positive feedback, wherein a gene upregulates its own expression.  Such an autoregulatory gene activator formed the basis of an experimental study \cite{hox1reversal} on the ``reversal of subfunctionalization'' in the pathway that governs brain-stem development.  While different developmental triggers activate the paralogous \emph{hoxa1, hoxb1} gene pair in modern mice, they were replaced by a single autoregulatory activator responsive to a common set of  with \emph{cis}-regulatory sites \emph{cis}-regulatory inputs, and the resulting organism was viable \cite{hox1reversal}.   It is this circuit of a self-regulating activator, the smallest unit of positive feedback, that is the object of our study.  

We study the consequences of duplication of this gene, both at the level of phenotype and in potential evolutionary outcomes, upon mutation.  When mutations appear in this circuit as per the subfuntionalization model, introducing complementary loss of function \emph{cis}-regulatory mutations that confers tissue-specific expression patterns \cite{force-genetics1999,force-subfunction05}, we find the emergence of a property of some developmental trajectories, namely redundancy of duplicates helps overcome mutational loss \cite{wang-redundancy96}.  The mutation of a gene expressed in a tissue is compensated for by the upregulation of its unexpressed partner in this model, called transcriptional backup \cite{kafri-pnas06}.  Mutations affecting the activating sites of the transcription factor give rise to the onset of oscillations, both stable and bursty.  Since the mathematical model for oscillations is also applicable to a diploid genotype with two alleles differing at their activating site, we examine the evolutionary fate of the duplicated gene haplotype in a diploid species via the fitness \cite{OttoYong2002,walsh2003,kondrashov-koonin04} of the oscillatory phenotype.  Fitness values are indexed by parameter values in the doubled activator circuit that give rise to distinct qualitative dynamics.

The dynamical system we study is that of a diploid model of the transcriptional network \cite{Omholt2000}, which enables us to consider both the interactions between diverging duplicates as well as allelic interactions \cite{Gjuvsland-PLOSone2010}, particularly those between allelic variants or heterozygotes.  Heterozygous advantage has been identified as a property that facilitates the fixation of a duplicate gene in a population \cite{spofford1969,OttoYong2002,proulx2006}, a finding of relevance to our results below.  Heterozygosity is also commonly associated with hybrid vigour; indeed, this correlation has also been extended to dosage (im-)balance of copy number variations and alterations to the amplitude of circadian rhythms \cite{hybrid-circadian2009}.   Unlike circadian clock models \cite{dunlap99}, which contain a negative feedback link \cite{novak-tyson-osc} in circuit topology, our duplicated activator network does not introduce negative feedback explicitly.  Instead, negative feedback arises due to a competitive mechanism whereby one activator excludes the binding of another to the promoter.  Although the mechanism implementing negative feedback -- and thus oscillations -- is different, our model displays a  dependence of amplitude on hetrozygosity and dosage balance found to be correlated to hybrid vigour \cite{hybrid-circadian2009}.   Furthermore, the presence of paralogous genes in oscillatory processes has been noted in the literature \cite{chen-WGDosc08,trachana-embo2010}, also as a means to maintain oscillations upon loss of single genes \cite{baggs-network09, shi-currbio2010}.  Although our model makes no explicit reference to the systems that contain the genes reported there, our theoretical finding suggests an evolutionary mechanism for the proliferation of duplicate genes that take part in oscillatory dynamics as heterozygotes.

There are two stages of modelling that we perform in this paper.  The key first step is to frame the onset of qualitative changes to the dynamics of transcription regulation as a consequence of gene duplication in the language of dynamical systems.  The second step is to model the likely evolutionary fate of a mutant haplotype containing such a duplicate.  For the first step, we set up a model of  transcriptional regulation of a genotype characterised by a duplicated positive feedback loop, with the transcription rates derived from the probability of promoter occupancy by RNA polymerase \cite{Ackers1982} in the presence of the activators.  An extension of the model to include the effects of intrinsic noise -- stochasticity in transcription --  is presented, to address how single cell and population averaged phenotype might differ.   To enable the second stage of analysis on the evolutionary fates, a population genetic model for the likely invasion of a duplicated gene is then introduced, with selection coefficients that depend on parameters of  the transcriptional network.   Thereafter, we present results on the behaviour of these models.  We identify qualitative shifts in the dynamics owing to alterations in  parameter values -- in \emph{cis} and in \emph{trans} --  to be presented separately.  After a discussion of the results owing to \emph{cis}-regulatory changes that affect switching behaviour, we address the case of changes that affect regulation by \emph{trans}-acting effects.  It is this set of changes that we shall track in the second stage of modelling, that of the evolutionary fate of the duplicated gene.   A final section summarises the different components discussed within the regulatory model, linking the qualitative aspects of model behaviour to different experimental studies.

\section{Model of the duplicated autoregulatory gene switch}
 \label{section:model}
 
\begin{figure}[htp] 
\centering 
\includegraphics[width=0.8\textwidth]{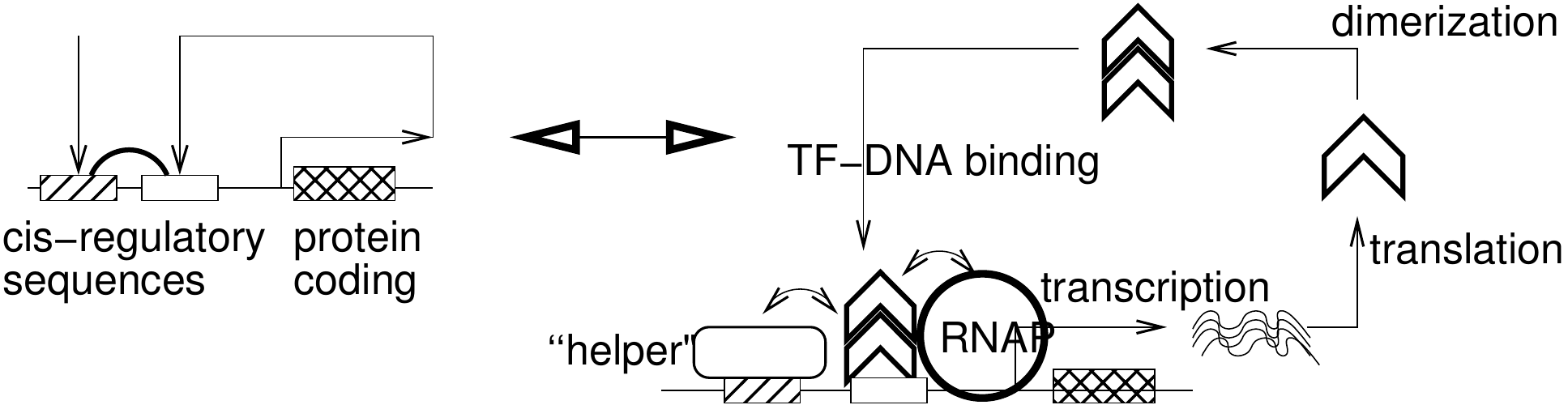} 
\caption{The haplotype on the left supports a representation of the schematic kinetics of transcription via dimeric activators which, with the adhesive reactions facilitated by the ``helper" proteins, bind to DNA and recruiting the RNA polymerases.  The mRNA transcribed is then translated, and homodimers are formed before the autoregulatory reactions proceed.  There are further reactions that involve decay of mRNA and proteins that are considered in the models below.}
\label{fig:grep}
\end{figure}

\begin{figure}[htbp]
   \centering
 \includegraphics[width=0.7\textwidth]{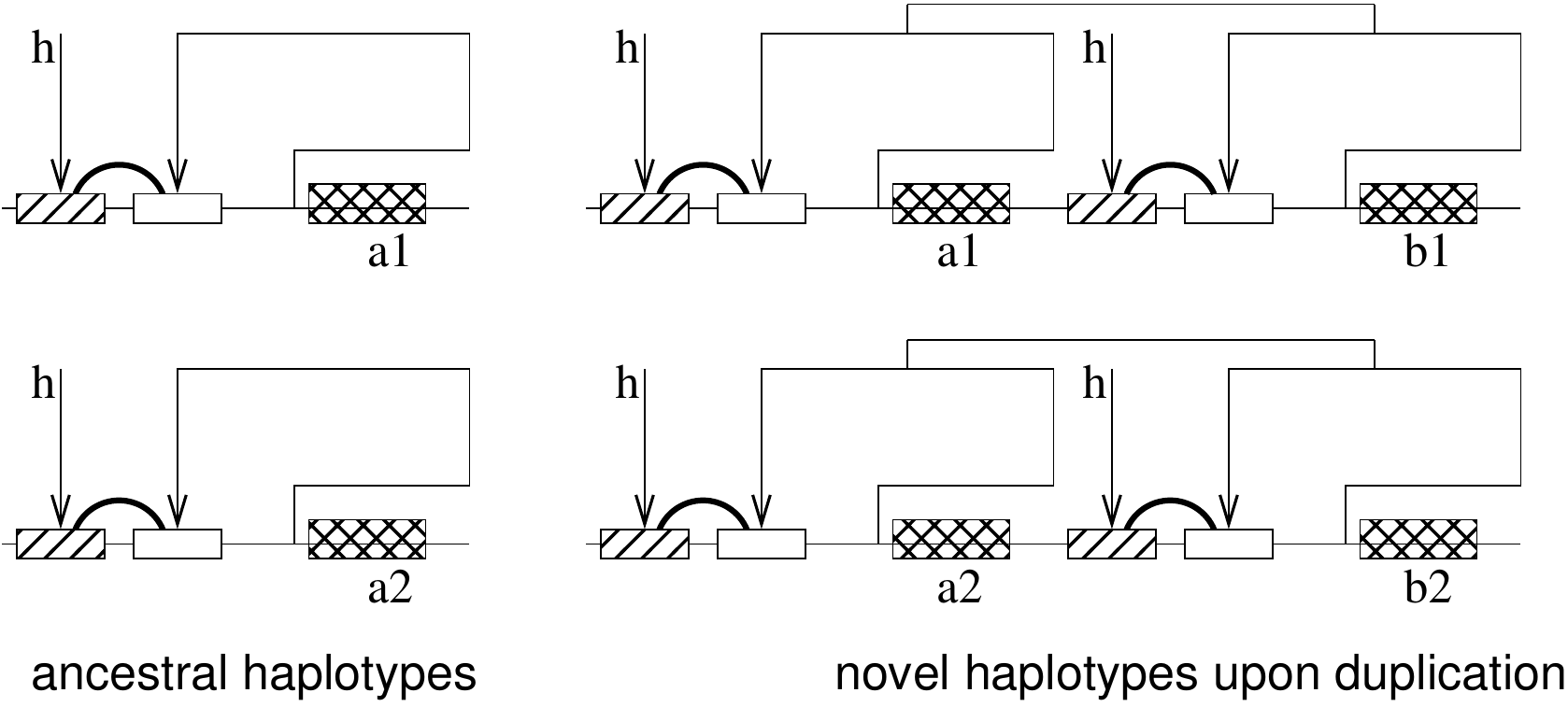}\\
 \includegraphics[width=0.7\textwidth]{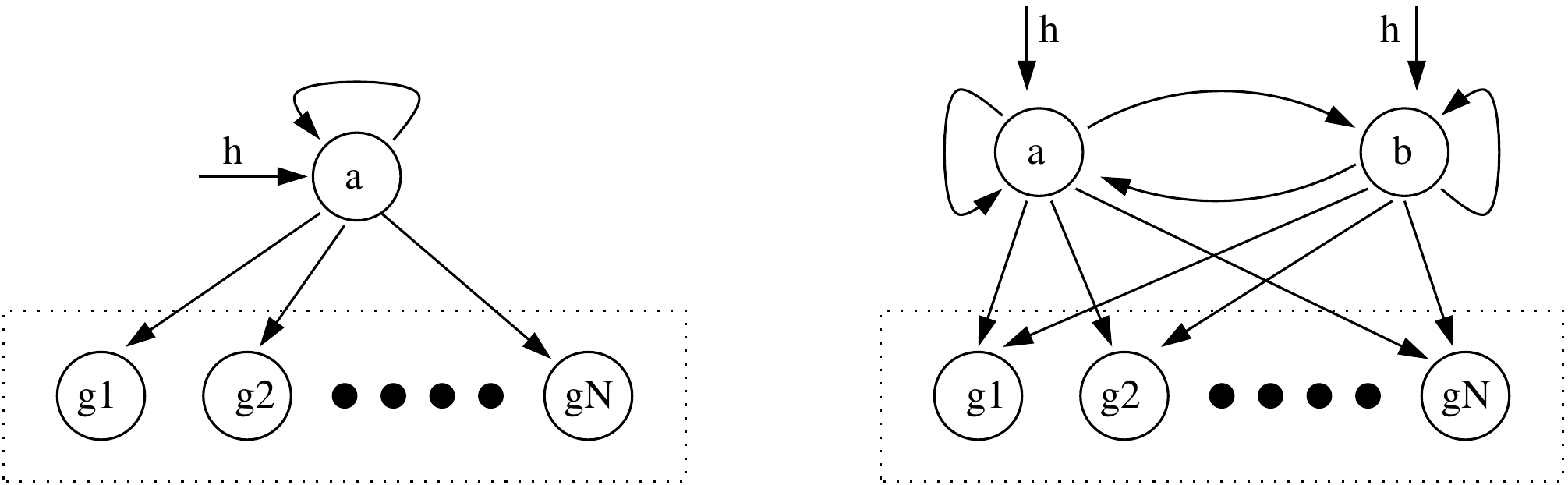} 
   \caption{{\bf Haplotype and network view of the duplicated gene.} The autoregulatory gene is part of the ancestral genotype and displayed in its network context on the left, with the allelic variants on top.  The duplicated haplotype leads to an increased number of regulatory interactions in the network on the right.}
   \label{fig:duplinet}
\end{figure}

A positive feedback loop provides a mechanism to convert transient input signals into stable output levels, acting as a switch.  Developmental stages, characterised by stable expression levels of subsets of genes, rely on regulatory circuits that implement positive feedback switches \cite{odom-hepatocyte2006,davidsonbook2006, hobert-terminalselectorPNAS08}.  The smallest circuit implementing positive feedback is one with a single autoregulatory gene (see Figure \ref{fig:grep}); for it to act as a switch,  it is necessary for the activation reaction of a transcription factor binding to the gene promoter to have a cooperativity index, or Hill coefficient, of 2 or more \cite{cherry2000}.   As a specific instantiation of such cooperativity, we require that the autoregulatory gene at the top of the hierarchy activates itself after dimerisation of its protein product (as shown in Figure \ref{fig:grep}), giving rise to a Hill coefficient of 2, although nothing in the model requires such a specific reaction.  In particular, we disregard consideration of heterodimeric or homodimeric association, and assume the existence of homodimers alone.  As will be clear from the analysis below, a greater degree of cooperativity facilitates many more qualitative changes to the dynamics, which we shall disregard in the interest of parsimony.  Our model takes such an autoregulatory  gene coupled to target genes, which makes up a topology of the ``terminal selector" network type \cite{hobert-terminalselectorPNAS08}, and duplicates it (see Figure \ref{fig:duplinet}).  When duplicating the gene, we shall assume that both coding and regulatory regions of the gene are duplicated; thus, we end up with the network on the right hand side of Figure \ref{fig:duplinet}.  Other influences on the activator might define developmental context or tissue specificity. Such interactions operationally outline the loci of context dependent changes that we shall introduce, and are indicated by the arrows labelled $h$ into the activators $a_{1,2}$ in Figure \ref{fig:duplinet}.

%

\subsection{Thermodynamic model of gene activation}
 
 The ``thermodynamic" approach to modelling transcription is to set the rate of transcript formation to be proportional to the occupancy of the promoter of the gene to be expressed from a single allelic locus \cite{Ackers1982}.  Promoter occupancy is facilitated by transcription factors that bind to cognate DNA sequences and recruit the transcription machinery -- RNA Pol II, mediator complex forming multi-component proteins, \emph{etc}.  The probability of occupancy is accounted for by assigning Boltzmann factors for the possible configurations of protein-DNA bound states \cite{Ackers1982,phillips2009physical}.  The binding and unbinding protein-DNA reactions of are assumed to be in rapid detailed balance to justify the use of the thermodynamic formalism; hence the ratios of these reaction probabilities are given by the negative exponential of the difference of free energies of the bound/unbound configurations.  
 
The contribution of these configurations to the promoter occupancy is modelled in detail in Appendix A.  Here we provide the resulting expression for the probability $\Phi_i$ of occupancy of the promoter of a gene $i$.  This can be introduced in terms of the function $\Psi_i$ of the transcription factors $\alpha_1, \alpha_2$ (represented in vector notation $\vect{\alpha}=(\alpha_1,\alpha_2)$)   
 \begin{equation}
\label{eq:foldchange}
\begin{array}{rcl}\vspace{10pt}
  \Psi_i(\vect{\alpha},
\vect{r}_i,\vect{t}_i) & = & \displaystyle\frac{ r_{i0}+ t_{i1}r_{i1}\alpha_1 + t_{i2}r_{i2}
  \alpha_2}{\displaystyle 1 + t_{i1}\alpha_1 + t_{i2} \alpha_2}
  \end{array}
\end{equation} 
which measures the amount of transcript produced in the presence of the $\alpha_i$ relative to the basal rate of transcription $r_{i0}$, when transcription factors are absent, a quantity called fold-change.  The parameters $r_{i1}, r_{i2}$ in $\mathbf{r}_i = (r_{i0},r_{i1}, r_{i2})$ stand for the strength of recruitment of the transcription machinery  (Pol II, Mediator, etc) \cite{ptashne-gann-book} by $\alpha_1$ (rate $r_{i1}$) and $\alpha_2$ (rate $r_{i2}$) from the chromosomal locus indexed by $i$.  The parameters $t_{ij}$ in $\mathbf{t}_i=(t_{i1}, t_{i2})$ stand for \emph{cis}-regulatory binding strengths of protein ($\alpha_j$) binding to DNA locus ($i$), determined by a more detailed description in eq. (\ref{eq:deftherm}) below.   The probability of promoter occupancy is given by 
 \begin{equation}
\label{eq:probpromoter}
\begin{array}{rcl}
\Phi_i(\vect{\alpha},
\vect{r}_i,\vect{t}_i) & = & \displaystyle\frac{1}{1+\Psi_i^{-1}(\vect{\alpha},\mathbf{t}_i,
\mathbf{r}_i)} \;\; (i=1,2) \mbox{ }.\vspace{10pt}
\end{array}
\end{equation} 

The derivation of these expressions for $\Phi_i$ and $\Psi_i$ follows \cite{Bintu2005a,phillips2009physical}, and is presented in  Appendix A.  There we also derive the expression \ref{eq:deftherm} below, which includes a ``helper" protein that serves as a proxy for contextual influences.  While the $A_j$ indicates the proteins of interest (the duplicated activators) the ``helper'' proteins $h$ are introduced to indicate the presence of co-regulators that enable tissue or developmental context-specific expression.   In the following expressions (\ref{eq:deftherm}), various $\varepsilon$s denote binding energies: $\varepsilon^{0}_{A d}$ denotes the non-specific binding of protein $A$ to any of $N_{ns}$ DNA binding sites; $\varepsilon^{s}_{A_k d_k}$ stands for the binding of $A_k$ to its cognate site $d_k$; $\varepsilon_{A B}$ measures the energy of protein-protein interactions between $A$ and $B$;  $\varepsilon_{hA_j p}$ the glue-like interaction between helper $h$, activator $A_j$ and the RNA polymerase/Mediator complex $p$:  
\begin{equation}
\label{eq:deftherm}
\begin{array}{rcl}
 \alpha_j &=&  \displaystyle  \frac{ A_j}{N_{ns}}, \\  
 r_{ij} &=&\displaystyle\exp(-\beta\varepsilon_{A_j p})\left(\frac{\displaystyle 1+\frac{ h}{N_{ns}}\exp(-\beta(\varepsilon^s_{h d_i}-\varepsilon^0_{h d}+\varepsilon_{h A_j p}-\varepsilon_{A_j p}))}{\displaystyle 1+\frac{ h}{N_{ns}}\exp(-\beta(\varepsilon^s_{h d_i}-\varepsilon^0_{h d}))}\right) \\
t_{ij}&=& \exp(-\beta(\varepsilon^s_{A_j d_i}-\varepsilon^s_{A_i d_i}))\displaystyle\left( 1+\frac{ h}{N_{ns}}\exp(-\beta(\varepsilon^s_{h d_i}-\varepsilon^0_{h d}))\right). \\   
\end{array}
\end{equation}
The details of the derivation is provided in \ref{context1}.

Eq. (\ref{eq:deftherm}) parameterises are the molecular interactions that determine transcription rates in the gene regulatory network.  While most of the paper will treat the $r_{ij}$, $t_{ij}$ (and other rates to be introduced shortly) as the parameters that determine network behaviour, the explicit definitions (\ref{eq:deftherm}) reveal the substrates upon which mutations may act in order to change the protein-DNA and protein-protein interactions that determine phenotypic outcomes and evolutionary fates.   In particular, in order to to present verbal arguments that rely on mutations that disrupt pleiotropic properties and restrict them to context-specific roles (as in the subfunctionalization model of \cite{force-genetics1999}; see also \cite{bergthorsson2007}) the helper protein index $h$ will serve as a placeholder for such context-specific factors.  This variable will be inherited by both copies of the gene upon duplication, and contribute to the set of mutations in \emph{cis} (labelled by $i$) by altering binding energy terms carrying both $i$ and $h$ indices.  Having multiple helpers allow \emph{cis}-context dependent effects by altering their binding affinities to DNA, and in particular in complementary ways -- one helper (and not the other) binds to locus 1 and not to locus 2  and vice versa.  We shall also study the effects of mutations that affect sites recruiting the transcriptional machinery and bring about changes in \emph{trans}-acting terms $\varepsilon_{A_j p}$ and/or $\varepsilon_{h A_j p}$,  with the case of multiple helper proteins detailed in \ref{context2}.  Measuring binding energies relative to $\beta = k_B T\sim 0.6$ kcal/mol, one can estimate the scale of the changes to parameters $(t_{ij}$, $r_{ij})\rightarrow (t'_{ij}$, $r'_{ij})$.  In particular,  a change in binding free energy of $\Delta\Delta G=1.5$ and $2.5$  kcal/mol, typical of biophysical measurements for mutant effects, translates to ratios of $t_{ij}/t'_{ij}$ or $r_{ij}/r'_{ij}$ of $\sim 12$ or $65$ respectively.  These are the parameter ranges for which qualitative changes to transcriptional dynamics are noted below.  The actual parameters used to perform the computational experiments as in \ref{parameters}.

\subsection{Model of haploid transcriptional deterministic kinetics}

In the thermodynamic model for the rate of transcript production, we make the assumption that the transcriptional activators $A_i$ act as homodimers, as shown in the representation of the autoregulatory circuit in Figure \ref{fig:grep} that we shall take to be the ancestral haplotype. Dimer formation by monomer binding and unbinding is assumed to rapidly be in detailed balance on the time scale of gene expression, so that the concentrations of dimers $[A_i]=x_i^2/K_{dim, i}$ where $x_i$ is the concentration of monomers and $K_{dim, i}$ is the dissociation constant for dimerisation.  Monomers are translated from transcripts which decay at rates $\delta_i$ much greater than the decay rates $\Delta_i$ of the corresponding proteins.  This assumption of faster time-scale of mRNA dynamics leads to the consequence that mRNA levels are slaved to protein dynamics.  The equations for the rates of changes to proteins $x_1$ and $x_2$  thus captures the essential dynamics of the system.  Here we present the model for the haploid case, with the detailed derivation provided in \ref{kinetic}: 
\begin{equation}
\label{eq:ODEmodel}
\begin{array}{rcl}
\dot{x_i}&=&c_i\varphi_i(\mathbf{x}, \mathbf{r}_i,\mathbf{t}_i) -
\Delta_i x_i, \quad (i=1,2),\\
\dot{\vect{y}}&=&\mathbf{f}(x_1,x_2,\vect{y})
\end{array}
\end{equation}
where $\mathbf{f}(x,\vect{y})$ is the dynamical sub-system for the set of downstream variables $\vect{y}$ that the activator gene $x$ influences, $\varphi_i(\mathbf{x}, \mathbf{r}_i,\mathbf{t}_i)=\Phi_i(\mathbf{x^2}, \mathbf{r}_i,\mathbf{t}_i)$ in eq. (\ref{eq:probpromoter}), with $\Delta_i$ the linear degradation rates for the proteins.
The variables $x_i$ have been scaled in terms of protein-DNA and dimer dissociation constants (see \ref{kinetic}), and the parameters $c_i$,  
\begin{equation}
\label{eq:cidef}
c_i=\displaystyle\frac{[\phi_i]\pi_i\mu_i}{\delta_i}
\end{equation}
are defined in terms of the rates for translation ($\pi_i$) and mRNA degradation ($\delta_i$).  The transcription rate $\mu_i$ expresses the proportionality between promoter occupancy in the thermodynamic description at each chromosomal locus, and $[\phi_i]$ captures the number of such loci, which will play an important role in our later discussion on haploid and diploid cases.

\subsection{Stochastic kinetic model}

The development of novel experimental techniques for tracking gene expression in single cells has made opened up for observation the consequences of the stochastic nature of the dynamics of gene regulation on phenotypes and their evolution\cite{raj-oud08,eldar2010,WangZhangNoiseFitness2011}.  In this section we present a simplified model of gene expression in this doubled autoregulatory gene network.  This will enable us to explore the consequences of intrinsic regulatory noise on the phenotype and what its implications might be for evolutionary fates of the duplicate genes.  While the detailed model of stochastic kinetics is provided in \ref{kinetic}, here we present a simpler model that incorporates many of the reaction steps into a Hill-type gene regulatory function, just as in the deterministic version in eq. (\ref{eq:ODEmodel}).  The reactions are summarised in Table \ref{tab:stochredreactions}.

\begin{table}
\caption{Reaction scheme for stochastic expression of proteins $A_i$, $i=1,2$.  There are 3 ways in which they are produced, by basal transcription, by regulated recruitment of transcriptional machinery by $A_1$ or by $A_2$ followed by translation.  In the above,  $\Gamma_i=(1+r_{i0})+(1+r_{i1})t_{i1}(\frac{A_1}{\kappa_1\Omega})^2+(1+r_{i2})t_{i2}(\frac{A_2}{\kappa_2\Omega})^2$, $\Omega$ is the volume factor and 
$\kappa_i$ is the geometric mean of the dissociation constants for dimerisation ($K_{dim}^i$) of protein $A_i$  and for protein-DNA binding $\kappa_i^2=K_{dim}^i \exp(\beta(\varepsilon^s_{A_i d_i}-\varepsilon^0_{A_i d}))$.}
\[
 \begin{array}{|c|l|}
\hline
 \mbox{reactions}& \mbox{rate of reaction} \\
\hline
\mbox{protein production:}&   \\
\phi_i\overset{\mbox{\tiny basal}}\longrightarrow A_i&\displaystyle c_i \frac{r_{i0}}{\Gamma_i\Omega}\\
\phi_i\overset{A_j}\longrightarrow A_i&\displaystyle c_i
\frac{r_{ij}}{t_{ij}\Gamma_i\Omega}\left(\frac{A_j}{\kappa_j\Omega}\right)^2\\\hline
\mbox{protein degradation:} & \\
A_i\longrightarrow \emptyset & \Delta_i\\ \hline
 \end{array}
 \]
 \label{tab:stochredreactions}
\end{table}

\subsection{Diploid model of regulation}

In diploid organisms, each gene comes in two copies independent of any duplication event.  To examine how duplication generates novelty, or to track how a genotype containing a mutant duplicate can be subject to evolutionary modelling, we need to construct a model of transcriptional regulation that can track allelic variants.  Further, we restrict differences between alleles to be solely in their activation sites, with different alleles (both before and after duplication) affecting transcription rates only by differing affinities for the transcriptional machinery -- $\varepsilon_{A_1 p}\neq \varepsilon_{A_2 p}$ and $\varepsilon_{h A_1 p}\neq \varepsilon_{h A_2 p}$.  Consequently the coefficients in $\mathbf{r}_i=(r_{i0},r_{i1},r_{i2})$ are independent of $i$.  To restrict our attention to activation site changes, based on results obtained from model analysis below, we shall assume that there are no differences between the alleles in the free energies of protein-DNA binding, hence we take $\mathbf{t}_i=\mathbf{t}_j=1$, where we have rescaled the concentration variable to be multiples of the protein-DNA and dimer dissociation constants that make up $\mathbf{t}_i$ (see \ref{kinetic}).  While the dynamics of transcriptional regulation will be studied in both $i$-dependent and $i$-independent cases below (where $i$ is the genomic locus), the evolutionary dynamics of duplicates will be explored only in the context of mutational variations affecting activating sites in a modular fashion, independent of \emph{cis}-regulatory context.

\subsubsection{Singleton case}

\begin{figure}[ht]
  \begin{center}
\includegraphics[width=0.85\textwidth,angle=0]{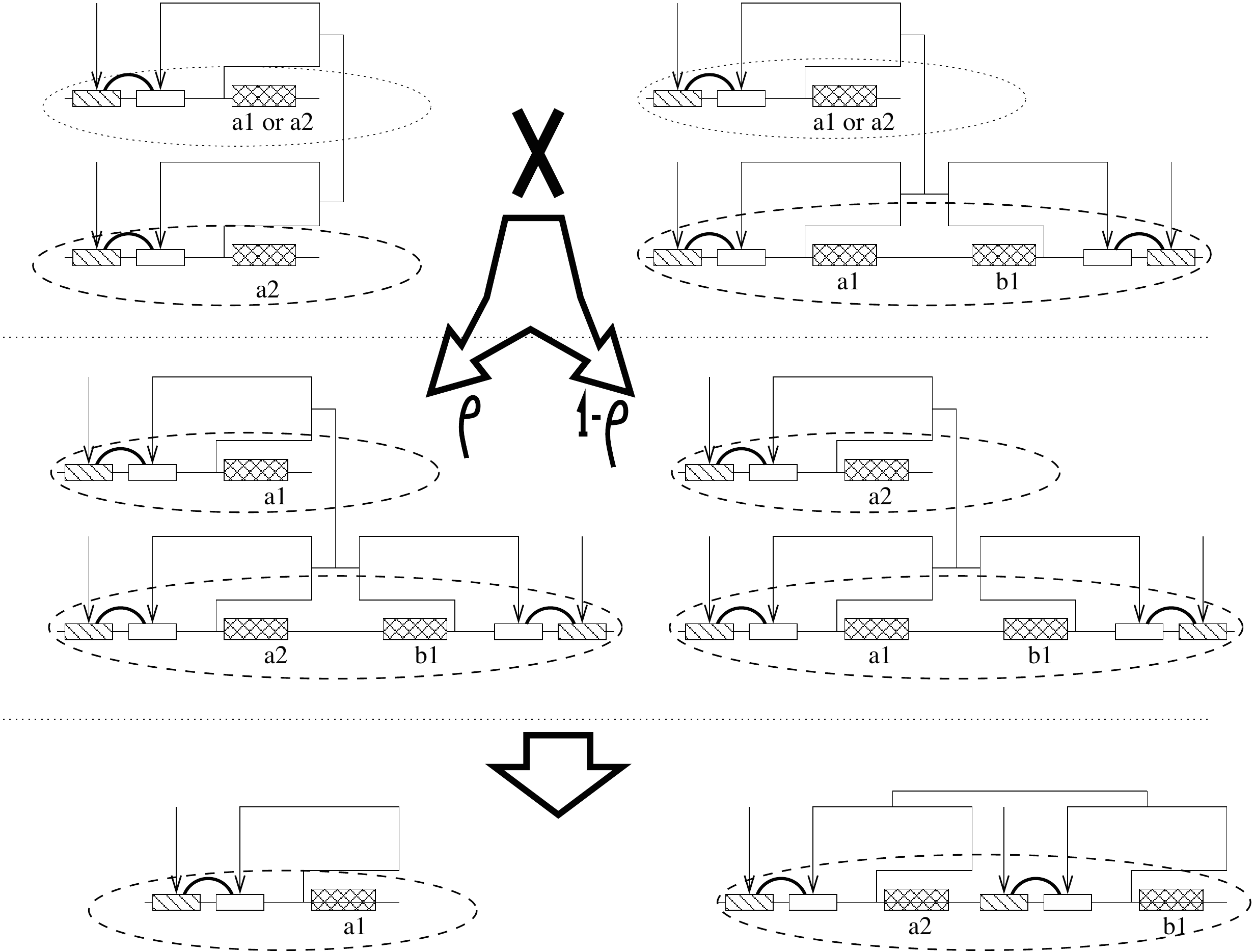}
 \end{center}
 \caption{{\bf Recombination amongst haplotypes and topologies for diploid regulatory networks.} Mating between two diploid genotype networks, one containing a chromosome with a duplicated segment of DNA, is shown with an X. The diploid network with duplicate genes on both chromosomes is not shown as that occur with a very small probability immediately after the duplication event.   A duplicate segment of DNA gets shuffled by the process of recombination, which is assumed to occur at a rate $\rho$.  Of the mating genotypes, the alleles chosen in the gametes are the dotted and dashed pairs, of which the dashed pair is tracked  in the figure as it undergoes recombination events. In the model presented below, singleton alleles are at their equilibrium frequencies in the population.}  
\label{fig:diploid}
 \end{figure}

If there are two allelic forms $x_i^{(j)}$ where $j=1,2$ for each gene $x_i$ we can extend the development leading to (\ref{eq:ODEmodel}) to consider how each allele generates a transcript at the rate determined by its promoter occupancy by transcription factors.  We can now set up a model to investigate variability in transcription dynamics owing to allelic variations, or heterozygosity for the autoregulatory circuit prior to duplication.  We introduce an allelic index $j=1,2$ in the fold-change $\Psi^{(j)}_i$ for each gene labelled by $i$ \cite{Omholt2000}.  For the case of a single locus $i=1$ with the corresponding fold change $\Psi^{(j)}_1(x_1^{(1)},x_1^{(2)})$.   If the two alleles $x_1^{(1)}, x_1^{(2)}$ are the same, the homozygous case, the diploid genotype network is shown in Figure \ref{fig:diploid} (A), but the model dynamical model is equivalent to the ancestral haplotype in Figure \ref{fig:duplinet}(a), or the switch model of a single autoregulatory gene, but with the parameter representing protein-DNA binding doubled: $t \mapsto 2 t$ (eq. (\ref{eq:foldchange}).  For the heterozygous case, allelic variants $x_1^{(1)}, x_1^{(2)}$ may be regarded as the two distinct activators $A_1$ and $A_2$ in the haplotype depicted in Figure \ref{fig:duplinet}(b).   The dynamical equations are then the same as in the haploid case, except that $x_1^{(1)}\mapsto x_1$ and $x_1^{(2)}\mapsto x_2$ in eq. (\ref{eq:ODEmodel}).

\subsubsection{Duplicated allele}
 
There are two cases to consider when one of the alleles is duplicated -- if the ancestral haplotype is homozygous or heterozygous.  If homozygous,  the duplicated genotype is again of the original haplotype topology as in Figure \ref{fig:duplinet}(a).  There is thus only one species of transcription factor protein which regulates itself and its downstream components $y$, with the effective dissociation constant to its DNA binding site scaled by a factor of 4 relative to the single copy haploid model.  This will only shift the the threshold for switching in the analysis presented below.

If the ancestral genotype is heterozygous, containing two different alleles $a_1$ and $a_2$ at a locus labelled by $A$, duplication creates another chromosomal locus, $B$, in a mutant haplotype containing 2 copies of one genetic sequence, say $a_1 b_1$.  The corresponding genotypes can in a mixed mating population, considered in greater detail below, are of the form $a_i/a_j$ (4 combinations for $i,j=1,2$) $a_i b_j/a_k$ (8 combinations) and  $a_i b_j/a_k b_l$ (16 combinations).  All these combinations require only 2 different species of transcription factors.  If the allelic variant is in a coding region, the differences show up in the rates of recruitment of the transcription machinery, $\varepsilon_{A_j p}$, whereas if the mutations are in the regulatory region, the differences to be considered are in the binding of helper/coactivator proteins, that change $\varepsilon_{h A_j p}$ in eq. (\ref{eq:deftherm}). 
(The inability to form heterodimers between distinguished alleles is still assumed.)  Upon taking these changes into account, the difference in the number of distinct gene copies to be considered is contained in the parameter $[\phi_i^{(k)}]$ that provides allelic specificity to $c_i$ in eq. (\ref{eq:cidef}).  

In sum, the corresponding model is described by the ODEs
\begin{equation}
\begin{array}{rcl}
\dot{x_i}&=&\sum_{k}c^{(k)}_i\varphi^{(k)}_i(\mathbf{x}, \mathbf{r}^{(k)}_i,\mathbf{t}^{(k)}_i) -
\Delta_i x_i, \quad (i,k=1,2),\\
&=&\sum_{k}c^{(k)}_i\varphi^{(k)}_i(\mathbf{x}, \mathbf{r}^{(k)},\mathbf{t}^{(k)}) -
\Delta_i x_i, 
\end{array}
\end{equation}
where (again) the superscript ${(k)}$ refers to the two alleles and their corresponding parameters refer to possibly different binding interactions at each allele as introduced.  The parameters $c_i$ in (eq. (\ref{eq:cidef}) can be extended to reflect potential allelic differences in transcription, translation or degradation rates for proteins and transcripts, but which we keep the same for both copies of the duplicated allele, since our interest lies in the mutation that leads to an allele with two copies.  Thus, to present the results in the latter sections, it is the model in eq. (\ref{eq:ODEmodel}) that will be studied with the parameter $c_i$ playing a role in the evolutionary discussion.

\subsection{Adaptive dynamics of duplicated gene}
\label{adaptive}

Most of what we have modelled  of the autoregulatory switch has been independent of the mode of its inheritance, to which we have appealed to in order to set up the terms of discussion, but which we have not  presented in detail to pursue its possible evolutionary implications.  In particular, our discussion has implied a continuity of argument from haplotype to inheritance and evolutionary fate.  As noted, the key difference introduced by duplication at the level of haplotype is the multiple feedback structure, an interaction topology that is novel for haploids but not diploids.  For the diploid case, the locus of change for system dynamics upon duplication is tracked by the parameter $c_i$ which incorporates the copy number $[\phi_i]$.  The number of functional copies of the gene used for Mendelian arguments and for matching their effects to the environments inhabited by the phenotype ranges from 0 to 4 in the diploid case \cite{kondrashov-koonin04}.  Recombination in sexually reproducing diploid organisms ``randomises'' novel mutations along genomes and may offer greater possibilities for fixation by selective forces \cite{hill-robertson66}.  In the previous section we have associated  the analysis of divergent properties of post-fixation duplicates to pre-duplication heterozygotes \cite{spofford1969,proulx2006} by mapping them onto a common dynamical system for transcriptional dynamics.  In this section, we investigate the evolutionary consequences of network level effects of homozygous and heterozygous genotypes using a model for the evolution of duplication based on \cite{OttoYong2002}.  In this model, a mutant duplicated allele is introduced into, and examined for its ability to invade, a population characterised by an existing equilibrium gene pool.  Equilibrium configurations provide standard population genetic backgrounds in which a mutant's ability to survive is analysed.  The transcriptional network of interest in this paper makes it natural to consider the effects of distinct alleles in the equilibrium population; hence, we examine the case of equilibrium maintained by overdominance, where heterozygotes are fitter than either homozygote in this bi-allelic setting.  We also do not wish to consider the effects of mutation after the duplication event in the evolutionary picture (as it is already a lengthy paper); thus, we eliminate consideration of equilibria involving mutation-selection and mutation-drift balance \cite{hartl-clark07}.  

The model includes haplotypes defined via a pair of loci $A$, $B$, with allelic values of $a_1, a_2$ for $A$ and $b_0, b_1, b_2$ for $B$.  $b_0$ is the null allele, so the haplotypes $a_1b_0$ and $a_2b_0$ are the alleles $a_1, a_2$ prior to duplication.  (Note, that we have introduced the null allele at the second locus for convenience.  If we allowed for a null allele, $a_0$ at the first locus, the pre-duplication alleles would, equivalently, be expressed as $a_0b_1$ and $a_0b_2$.)  In the overdominant case, the model assumes an initial polymorphic equilibrium configuration of  alleles $a_1$ and $a_2$, via heterozygote advantage \cite{spofford1969}.  These singleton haplotypes are also called $a_1b_0$ and $a_2b_0$  below.  Homozygotes $a_1/a_1$ and $a_2/a_2$ (also named $a_1b_0/a_1b_0$ and $a_2b_0/a_2b_0$) have fitness coefficients $1-s$ and $1-t$ relative to the heterozygote $a_1/a_2$ with a  fitness coefficient of 1; consequently, the relative frequencies of alleles $a_1$ and $a_2$ at equilibrium are $\hat{x}_{10}=t/(t+s)$ and $\hat{x}_{20}=s/(t+s)$, respectively \cite{hartl-clark07}.  The fitness of the population at this polymorphic equilibrium is $\hat{W}=1-ts/(s+t)$.  

A duplicate mutant is introduced into a background of this existing equilibrium condition,   The duplicate haplotypes are denoted $a_1b_1, a_2b_1, a_1b_2, a_2b_2$ with frequencies $x_{11}, x_{21},x_{12},x_{22}$ respectively.     Following \cite{OttoYong2002}, we assume a two step decomposition for updating the haplotype frequencies:
\[
 x_{ij}\mapsto x_{ij}^* \mbox{ by recombination, and we shall ignore }  x_{ij}^*\mapsto x_{ij}' \mbox{ by mutations.}
 \]
 We set up the discrete dynamics of frequency updates due to recombination by first assuming that the mutant duplicate has a probability of being chosen for mating with a probability $\sim N_e^{-1}$ where $N_e$ is the effective population size, much smaller than those of the singletons $\hat{x}_{10}=t/(t+s)$ and $\hat{x}_{20}=s/(t+s)$ which are at their equilibrium frequencies.  Therefore the probability of pairing duplicate gametes with each other is negligible compared with those of pairing a duplicate haplotype with that of a singleton \cite{OttoYong2002}.  After duplicating $a_1$ to produce $a_1 b_1$,  recombination, mating individuals containing gametes of haplotype $a_1b_1$ with those containing $a_2b_0$ occurs with probability proportional to $\hat{x}_{20} x_{11}$,  creating a $a_1b_0/a_2b_1$ genotype at a rate proportional to recombination probability $\rho$ and $a_2b_0/a_1b_1$ with probability $(1-\rho)$.  The fitness values of the two genotypes are indexed by their subscripts -- $W_{10;21}$ and $W_{20;11}$ respectively.  To first order, the equilibrium frequencies $\hat{x}_{10}$ and $\hat{x}_{20}$ are unchanged.
 
 The discrete map is represented via the block-diagonal matrix: 
\begin{equation}
\label{eq:adaptive}
\begin{array}{lcr}
\left(\begin{array}{c}
x^*_{11}\\x^*_{21}\\x^*_{22}\\x^*_{21}
\end{array}\right)&=&
\displaystyle \frac{1}{\hat{W}}\left(
\begin{array}{lr}
\mathbf{R_{11\times21}}&\mathbf{0}\\\mathbf{0}&\mathbf{R_{22\times12}}
\end{array}
\right)
\left(\begin{array}{c}
x_{11}\\x_{21}\\x_{22}\\x_{12}
\end{array}
\right),
\end{array}
\end{equation}
where the $(2\times 2$ submatrices are defined in (\ref{eq:admatrices}) below.
\begin{equation}
\label{eq:admatrices}
\begin{array}{ll}
\mathbf{R_{11\times21}}&=
\left(
\begin{array}{lr}
W_{10;11} \hat{x}_{10} + (1-\rho) W_{20;11}\hat{x}_{20} & \rho W_{10;21}\hat{x}_{10}\\
\rho W_{20;11}\hat{x}_{20}&W_{20;21} \hat{x}_{20} + (1-\rho) W_{10;21}\hat{x}_{10} 
\end{array}
\right), \\&\\
\mathbf{R_{22\times21}}&=
\left(
\begin{array}{lr}
W_{20;22} \hat{x}_{20} + (1-\rho) W_{10;22}\hat{x}_{10}  & \rho W_{20;12}\hat{x}_{20}\\
\rho W_{10;22}\hat{x}_{10}&W_{10;12} \hat{x}_{10} + (1-\rho) W_{20;12}\hat{x}_{20}  \\
\end{array}
\right)
\end{array}
\end{equation}

The fitness allocation for $a_i b_0/a_i b_0$ is taken to be the same as that of the $a_i b_i/a_i b_0$ genotype, for $i=1,2$, so $W_{10;10}=W_{10;11}=1-s$ and $W_{20;20}=W_{20;22}=1-t$.  This assumes that, since identical genes can only generate switch-like behaviours, the outcome of increasing dosage is a shift in the threshold for switch activation only, which we assume to be neutral (see below).  For the case where the copy numbers for the allelic variants of the single genes are in the ratio $2:1$ or $1:2$, we take $W_{10;12}=W_{10;21}=W_{20;11}=1+d$ and $W_{10;22}=W_{20;12}=W_{20;21}=1-u$.  We shall make a note of the case $d=0$ since we would like to impose fewer constraints for positive selection explicitly.   All of these fitness coefficients are normalised with respect to the pre-duplication heterozygote, $W_{10;20}=1$.  The fitness coefficients are summarised in  Table \ref{tab:tablefitness}.

\begin{table}
\caption{The fitness values $W_{ij;kl}$ for the different genotypes obtained by mating $ij$ with $kl$ gametes.  The selection coefficients are with reference to the singleton heterozygote which is assigned a fitness of $1$.  All  coefficients $s,t,u$ are positive, indicating reduced fitness compared to the $a_1/a_2$ heterozygous singleton, while both signs of $d$ are briefly explored in the main text. 
}
\[
\label{tab:tablefitness}
\begin{array}{| c | c c c c c c |}
 \hline    
W_{ij;kl} & 10 & 20 & 11 & 21 & 22 & 12\\
  \hline    
10 & 1-s & 1 & 1-s & 1+d  & 1-u & 1+d\\   
20 & 1 & 1-t & 1+d & 1-u  & 1-t & 1-u\\                     
  \hline  
\end{array}
\]
\end{table}

Following standard practice, we have summarised the effects of viability and reproductive success by single scalar-valued parameters.  We will need to relate these parameters to dynamical states in the model of the autoregulatory gene activator in order to make claims about evolutionary consequences of regulatory changes.  
Thus, $s$, $t$, $d$, $u$ will be defined in terms of the parameters in the reaction system shown in \ref{tab:reactions2} in the Appendix, and which for the purpose of dynamical analysis we have summarised in terms of the parameter combinations that arise in (\ref{eq:ODEmodel}).  Hence the selection coefficients of genotypes $a_kb_l$ capture dependence on the network parameters $s=s(\mathbf{r}_i,\mathbf{t}_i,c_i,\Delta_i)$  in a specific environment via their dynamical behaviour.  Since we do not model any particular environment in this paper, we will require verbal arguments indicating the plausibility of adaptive roles of the dynamical states in parameter space. The passage from a continuous set of biochemical parameters to a discrete set of selection coefficients will be motivated via the appearance of distinguished qualitative regions in phase space (see Figure \ref{fig:osc-all} below) that partition the phenotype space into a discrete set of qualitatively different dynamical behaviour.  
%
%

\section{The two component subsystem feeds novelty downstream}

In this section we will argue that the autoregulatory component of the network in Figure \ref{fig:duplinet} can be analysed in isolation of its downstream effects for the kinds of effects that we will focus on -- the effects of dosage (im)balance introduced by gene duplication.  This will be obtained by showing how the eigenvalues of the linearised dynamics can be factorized.  We then show an immediate consequence of competition between the two copies of a gene for regulatory binding and activation.  

\subsection{Restriction to the two-component subsystem}

The Jacobian of the dynamical system in eq. (\ref{eq:ODEmodel}) has the same structure as the adjacency matrix in Figure \ref{fig:duplinet}).  Since the Jacobian determines the local dynamical behaviour around any state of the system, changes to its eigenvalues signal the onset of qualitative behavioural patterns.  In particular, around each fixed point of the dynamical systems of the gene regulatory networks shown in Figure \ref{fig:duplinet}, the Jacobian of the network gets updated as shown :
\begin{equation}
\left(
\begin{array}{c|ccc}
  a_{11} & 0 & 0 & 0 \\ \hline \\[-9pt]
  g_{11} & \multicolumn{3}{c}{\multirow{4}{*}{{\Huge $\mathcal{G}$}$_{\!\! N\times N}$}} \\[-4pt]
  \vdots & \\
  g_{N1} &
\end{array}
\right)
\longrightarrow
\left(
\begin{array}{cc|ccc}
  a_{11} & a_{12} & 0 & 0 & 0 \\ \\[-12pt]
  a_{12} & a_{22} & 0 & 0 & 0 \\ \hline \\[-12pt]
  g_{11} & g_{12} & \multicolumn{3}{c}{\multirow{4}{*}{{\Huge $\mathcal{G'}$}$_{\!\! N\times N}$}} \\[-4pt]
  \vdots & \vdots & \\
  g_{N1} & g_{N2} &
\end{array}
\right).
\label{eq:duplinet-jacobian}
\end{equation}
 In eq. (\ref{eq:duplinet-jacobian}), the $(\mathcal{G}_{N\times N})=\frac{\partial \mathbf{f}}{\partial \mathbf{y}}$ submatrix of the Jacobian corresponds to the part of the network enclosed in a dotted box in Figure \ref{fig:duplinet};  $g_{i1}$ ($i=1,\ldots,N$) are the partial derivatives of the promoter occupancy functions of the downstream genes with respect to the variable representing the autoregulatory gene.  $a_{11}$ denotes the partial derivative of the net rate of expression of the autoregulatory gene with respect to its own expression state.  After duplication (or in the biallelic case), there are two genes that influence the downstream sub-network, and auto- and cross-dependencies in the regulatory dynamics of the activators with Jacobian elements $a_{ij}$ are introduced.  The block structure of the Jacobian matrices implies that the determinants are of the form 
\begin{equation}\label{eq:factored-det}
a_{11}\times\vert\mathcal{G}\vert  \mbox{ and } \left\vert \begin{array}{cc} a_{11} & a_{12} \\  a_{21} & a_{22} \end{array}\right\vert \times \vert\mathcal{G'}\vert.
\end{equation}
and thus the eigenvalues that determine dynamical consequences factor into two pieces.  This modular decomposition suggests the following strategy.    

The argument invoking the selective advantage of increased gene dosage \cite{kondrashov-koonin04} looks to the effect of the doubling of steady state levels $x^*\rightarrow(x_1^*+x_2^*)$ on the target gene dynamics $\vect{f}(x^*,\vect{y})$ (where we use $^*$ to indicate steady-state levels in the regulatory network in a feed-forward manner.  These changes $\vect{f}(x_1^*+x_2^*,\vect{y})-\vect{f}(x^*,\vect{y})$ may indeed be associated with a positive selection coefficient. However, we shall instead look at other sources of qualitative shifts -- not gene dosage, but gene dosage balance instead.  We shall assume that the autoregulatory positive feedback serves to set a binary decision switch for the downstream components to be triggered, and the net effect of this doubling `merely' adjusts the threshold of the switch.  Since we look for signals of dynamical shifts via the local analysis provided by the eigenvalues of the Jacobian, and the Jacobian in eq. (\ref{eq:duplinet-jacobian}) is of a factorised form, the evolutionary significance of the changes ascribed to downstream sensitivity to doubled gene dosage lies in $\vert \mathcal{G'}\vert$ in eq. (\ref{eq:factored-det}). In this paper, we are interested in the qualitative changes that might influence phenotypes of duplicated genes lies in the other factor involving the $a_{ij}$ components.  In considering dynamical changes, such as the onset of oscillations to be considered below, this explains why we focus on the two-gene sub-network involving the duplicated autoregulator. For the case of the switch, we shall use the downstream levels as a reporter of the dynamical switches in the duplicated gene motif. 

The set of reductions required to get to this simplified form (\ref{eq:ODEmodel}) assumes separation of time-scales of transcript and protein formation and degradation relative to protein-DNA and protein-protein interactions.  This can introduce differences in the time scales of the results, particularly when dimerisation is involved \cite{bundschuh03slow} -- the full system has slower dynamics
which shows up in its time-dependent behaviour.  Similar consideration needs to be paid to the time scales of the interactions that couple the double-activator motif to the downstream components.  It is known that coupling to downstream components can alter qualitative dynamics of a network, a principle dubbed ``retroactivity" \cite{sontag-retroactivity}.  We have checked that the dynamics of the coupling can indeed affect the behaviour of the system, as does the kinetics of dimerisation.  However, we have checked that it is possible to find parameter ranges for downstream coupling and dimerisation kinetics for which the behaviour of the full system behaves in a manner qualitatively similar to the simplified model we choose to focus on, albeit with a slower time-scale for the dynamics.   However, for the purposes of this paper, these differences are not significant;  all the essential qualitative features predicted from this simple model are also present in the full kinetic description.

\begin{figure}[ht]
\centering
\includegraphics[width=0.8\textwidth, height=0.2\textwidth]{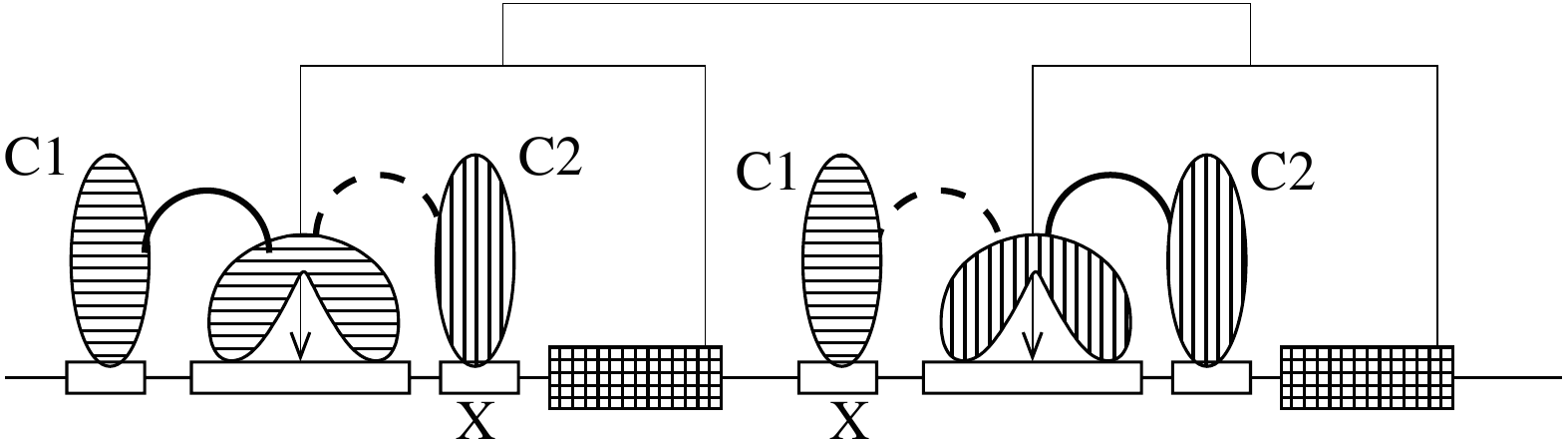}
 \caption{{\bf Binding strengths and \emph{cis}-regulatory mutation leads to the heterozygous switch.} The two proteins expressed by the two alleles/duplicate genes are shown in vertical and horizontal stripes bound to their DNA binding site indicated by the target of the regulatory arrow.  They have complementary binding strengths to the helper proteins $C_{1,2}$ appropriately shaded and further indicated by the arcs between the proteins -- the dark arcs denote greater strength of association compares to the dashed ones.  The sites marked X are where complementary loss of binding occurs upon mutation, as is postulated in the subfunctionalization model \cite{force-subfunction05}. }  
     \label{fig:cisreg}
\end{figure}

For the analysis below, we shall focus on two principal cases for the combinatorial regulatory parameters $\mathbf{t}_i,\mathbf{r}_i$.  In one, we set $\vect{r}_1=\vect{r}_2=\vect{r}$, corresponding to the case when recruitment for activation is modular, \emph{i.e.} independent of \emph{cis-} context $i$.  Thus,  $(r_{10},r_{11},r_{12})=(r_{20},r_{21},r_{22})=(r_0,r_1,r_2)$ and  $\varphi_1=\varphi_2=\varphi$.   When {\em cis}-regulatory context matters for expression, for instance, in the scenario described in the subfunctionalization model, we consider $\mathbf{r}_1=(r_0,r_1,r_2)$ and $\mathbf{r}_2=(r_0,r_2,r_1)$.  If $r_1>r_2$, this choice indicates a greater rate of transcription of gene 1 from duplicate locus 1 and a greater transcription rate for gene 2 from locus 2.  This can be achieved by assigning different affinities for the helper proteins $h$  in the two contexts labelled by $i$ in equation (\ref{eq:deftherm}).  A detailed derivation of how this emerges is provided in  \ref{context2}, and Figure \ref{fig:cisreg} makes the modelling assumptions explicit.

\subsection{Dual regulation as a consequence of changes to activation domain}
\label{sec:dualreg}

The fold-change function $\Psi_{\ell}$ (\ref{eq:probpromoter}) determines whether the influence of a protein on gene expression is that of an activator or a repressor, indicated by the sign of its derivative with respect to the amount of transcription factor.  The derivatives of $\Psi_\ell$ with respect to $\alpha_{1,2}$ is given by 
\begin{equation}\label{eq:dualreg}
\begin{array}{rcl}
  \left(
\begin{array}{c}
\displaystyle
\partial/\partial \alpha_1\\
\displaystyle
\partial/\partial \alpha_2
\end{array}
\right)
 \Psi_\ell(\vect{\alpha}, \vect{t}_\ell,\vect{r}_\ell)&=&\displaystyle\frac{t_{\ell1}t_{\ell2}(r_{\ell1}-r_{\ell2})}{(1+t_{\ell1}\alpha_1+t_{\ell2}\alpha_2)^2}\left(\begin{array}{c}\displaystyle
 \frac{r_{\ell1}-r_{\ell0}}{t_{\ell2}(r_{\ell1}-r_{\ell2})}+\alpha_2\\\displaystyle
 \frac{r_{\ell2}-r_{\ell0}}{t_{\ell1}(r_{\ell1}-r_{\ell2})}-\alpha_1
 \end{array}
 \right),
\end{array}
\end{equation}
flagging the possibility of non-monotonicity of $\Psi_\ell$.  This non-monotonicity means that increasing the amount of a transcription factor produces a fold change of transcription that increases (activates)
in one context and decreases (represses) in another, a feature called dual regulation.  In this model, the context is set by  the concentrations of the paralogous protein.  We consider the case $r_{\ell 1}>r_{\ell 2}$, where the $A_1$ binds more strongly to the Pol II enzyme than $A_2$ (ignoring enhancer context $\ell$ for $h=0$ in (\ref{eq:deftherm})) and find that $A_2$ behaves as an activator at low levels of $A_1$ 
and a repressor when the $A_1$ level $\alpha_1$ crosses a threshold \cite{vijay-thesis,dasmahapatra-wcsb2011}.  This repressor-like behaviour of $A_2$ (the activator with weaker affinity to Pol II) occurs even when $A_2$ is a facilitator of transcription by itself.  Since access to the binding site on the DNA is a limiting resource, with both activators competing for it, increasing the levels of the weaker activator hampers the overall efficiency of transcription from the combined $(A_1, A_2)$ system. 
Unequal binding affinity to DNA target sites ($t_{\ell 1}$,$t_{\ell 2}\neq1$) changes the amount of regulator 
$A_1$ that must be present for the crossover behaviour to occur.   Such dual regulation can also be a property of activators expressed from a single polymorphic locus.
%
%

This feature, that duplication of an autoregulatory gene can introduce competition for regulatory sites on the DNA and lead to dual regulation, opens up the possibility that novelty can arise due to mutations to modular components of one of the duplicated proteins which alters $r_{ij}=\exp(-\beta \varepsilon_{A_j p})=:r_j$ or $\exp(-\beta \varepsilon_{h A_j p})$ or even the DNA binding energy of the helper protein as seen in eq. (\ref{eq:deftherm}).  In all of what we discuss in this paper, the key parameters that we change are the $r_{ij}$ in eq. (\ref{eq:deftherm}) -- the arrow connecting transcription factor to the ``helper" in Figure \ref{fig:grep} -- while we keep $t_{ij}$ to be the same across the two genes of interest, just for simplicity.  The $r_{ij}$ do depend on protein-DNA binding strengths, but only for the helper proteins, as shown in \ref{context2}.

\section{Changes to dynamics of the network -- switches and backup}

The autoregulatory circuit behaves like a switch because the dynamical system (\ref{eq:ODEmodel}) supports 2 stable fixed points for a given choice of parameters.  Initial conditions specifying protein levels on either side of a threshold value drives the system to its low or high expression state.  In this section we consider the cases where having two genes coupled via feedback gives rise to different types of switches, referred to in \cite{guantes-poyatos-switch} as a progression switch (where two genes can move from low to high expression states) or a decision switch (where the genes switch to low-high or high-low expression states, the simplest example of a choice of regulatory fate).  

\subsection{Homozygous or progressive switch: lowering the on-off threshold}

\begin{figure}[ht]
\centering
\subfloat[]{
\includegraphics[width=0.4\textwidth, height=0.33\textwidth]{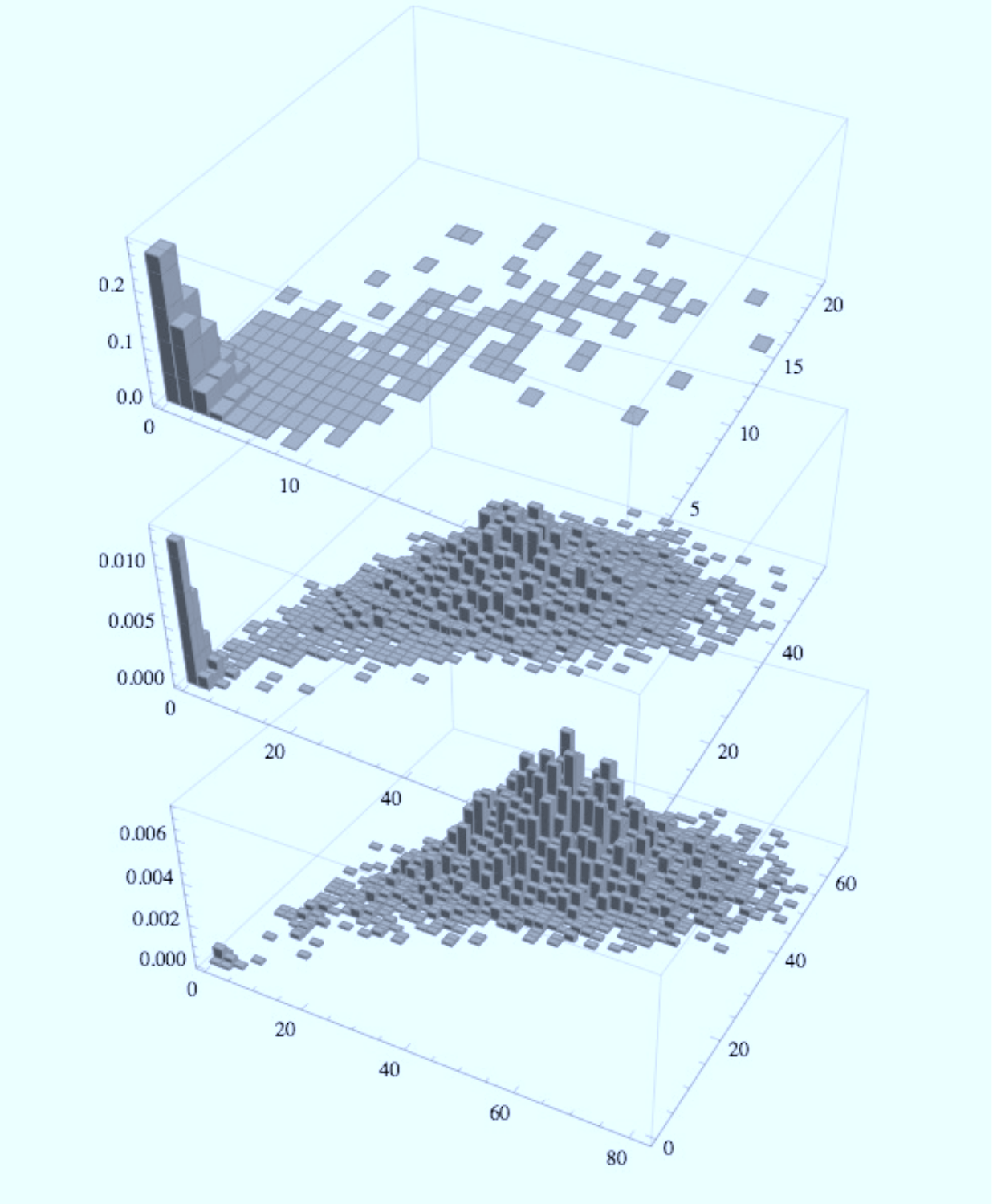}}
\subfloat[]{\includegraphics[width=0.33\textwidth]{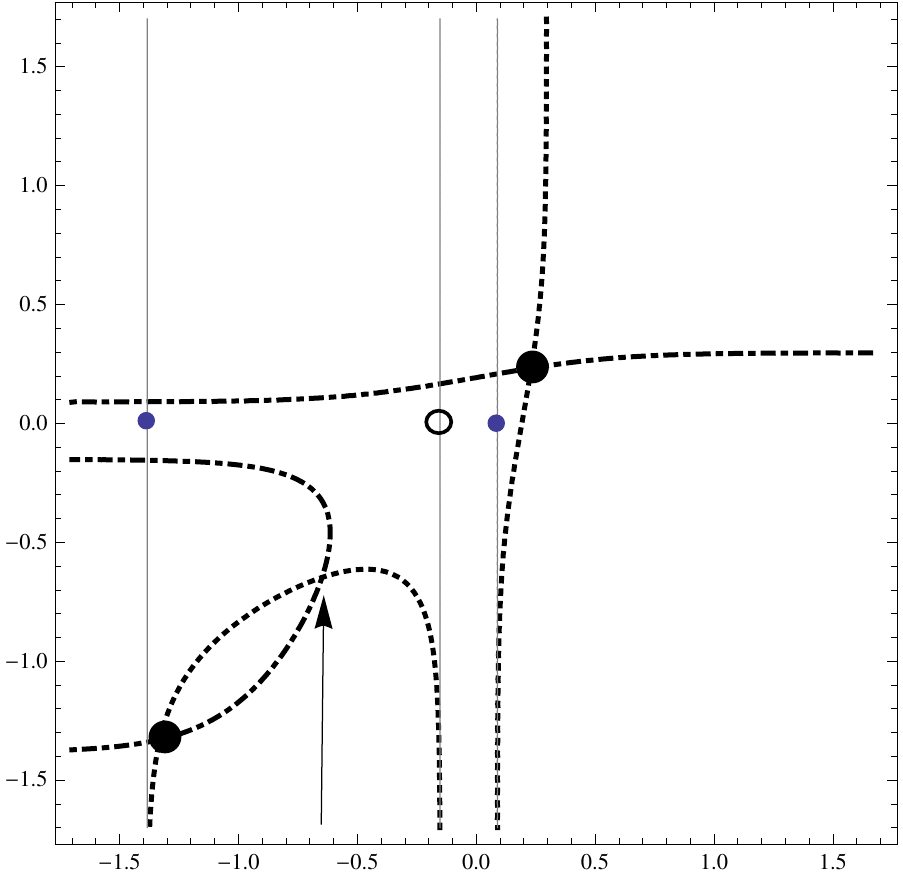}}
\subfloat[]{\includegraphics[width=0.3\textwidth]{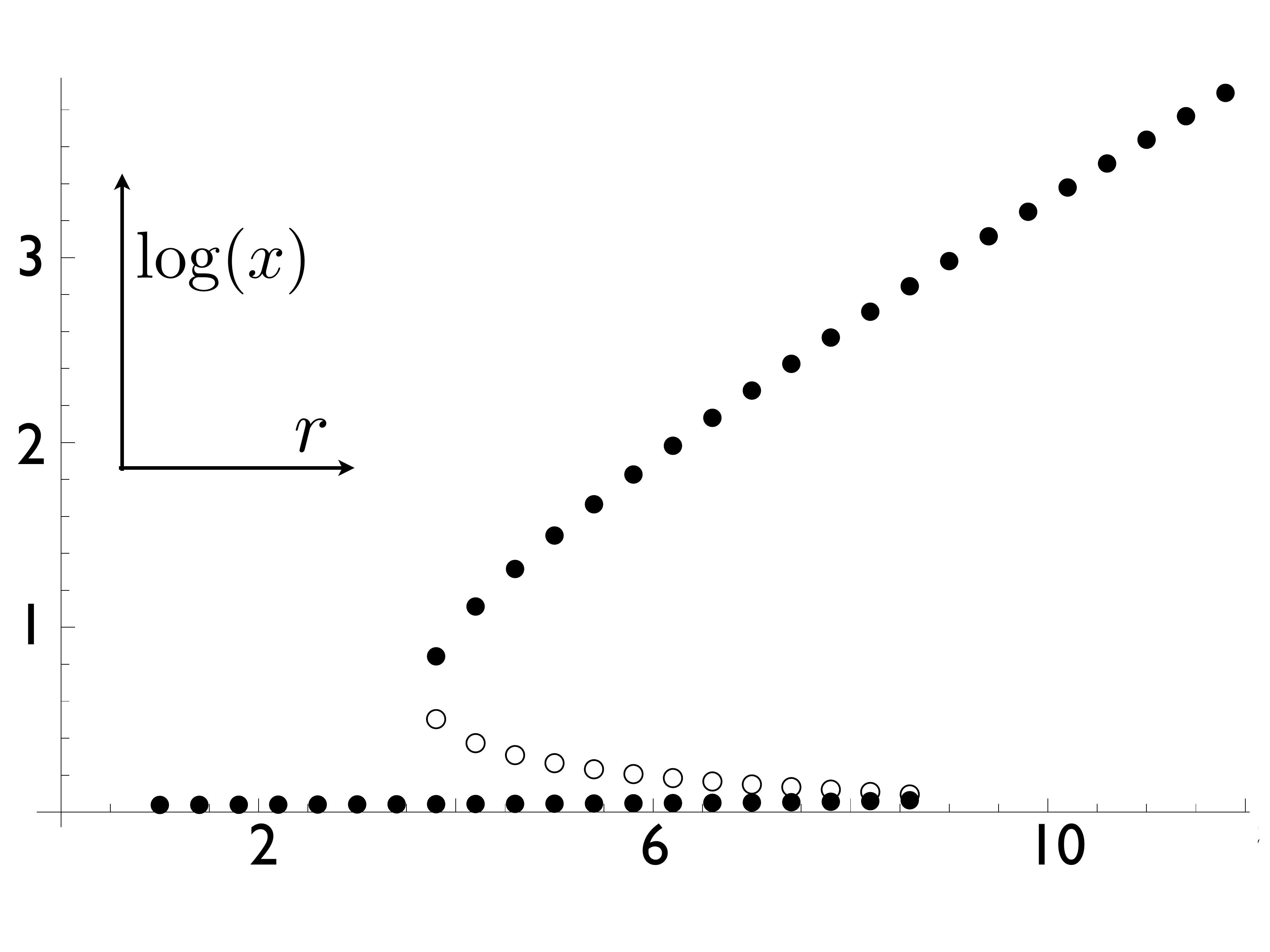}}
 \caption{{\bf The location of a transition from a low expression
     state to a high expression state is lowered upon duplication with no further divergence.} On the left are histograms of the expression levels of the duplicated self-activating gene, with the bifurcation from one to two to one stable steady state for changing $r_{ij}=r$ from 1 through 10 to 20 (from top to bottom).  All other parameter values are as in \ref{parameters},  with $c=\Delta=1$, and the activation strengths of the (identical) proteins taken as $r\times r_0$.The location of the modes (specifically shown for the histogram in the middle) correspond to the intersections of the (dotted, dot-dashed) nullclines indicated by the big black circles on the right.  The vertical lines in (b) correspond to the fixed points for the single autoregulatory loop, with the smaller dots indicating steady state levels there.  The open circle on the vertical line in the middle and the intersection indicated by the arrow are the locations of the thresholds in the single and duplicate gene cases respectively. In (c) the abscissa denotes increasing values of $r$ while the ordinate is the expression level.  The characteristic S-shape of a saddle node bifurcation with hysteresis is seen. The $x_1$-$x_2$ planes in this figure are in logarithmic scale.}  
     \label{fig:homswitch}
\end{figure}

Duplication of a gene has typically been associated with increase of protein product.  If the gene were part of a switching circuit with a phenotype that is binary-valued, as is the case for networks that convert transient, threshold-crossing inputs into stable outputs, it might well be the case that doubling of a gene does not greatly increase protein product, but merely makes the threshold more accessible for crossing.  We illustrate this possibility by setting $r_{ij}=r$ and $\Delta_i=\Delta=1$ and $c_i=c=1$ in (\ref{eq:ODEmodel}).   This is equivalent to a single gene switch but with $t_{ij}$ doubled.  As expected, altering the dissociation constant for an autoregulatory circuit changes the threshold in a sigmoidal function.  The steady states are solutions to a cubic equation with two stable fixed points separated by a saddle.  The bifurcation from a mono-stable to a bistable state is of a saddle-node variety \cite{strogatzbook}.   The nullclines, bifiurcation plots and histograms of expressed proteins generated from a Gillespie simulation make the point vividly in Figure \ref{fig:homswitch}.  This subsection is presented only to make the comparison with the asymmetric case obvious.

\subsection{Heterozygous or decision switch via complementary loss of recruitment}

In this case, we note that the two duplicate proteins recruit Pol II with the help of complementary helper proteins at the two loci. Gene copy 1 is activated by protein 1 at a rate that is greater than that achieved by duplicate protein 2, $r_{11}=r \times r_{12}$ with $r>1$.  The roles are reversed for expression from gene copy 2, with $r_{22}=r'\times r_{21}$ with $r'>1$.  The detailed origin of these parameters lies in the loss of complementary regulatory binding sites of two helper proteins that have complementary binding preference to proteins of the two alleles/genes, as explained in \ref{context2}.  They are summarised in the Figure \ref{fig:cisreg}. The combined strength of recruitment of the transcriptional machinery is partitioned (as shown in Figure \ref{fig:cisreg}, and in detail in \ref{context2}) via loss of binding sites for the helper proteins.  For simplicity, we shall take $r_{21}=r_{12}$ and $r=r'$; the symmetry does not introduce any non-generic features to this dynamics, but makes the analysis more transparent.

\begin{figure}[ht]
  \begin{center}
\includegraphics[scale=0.36]{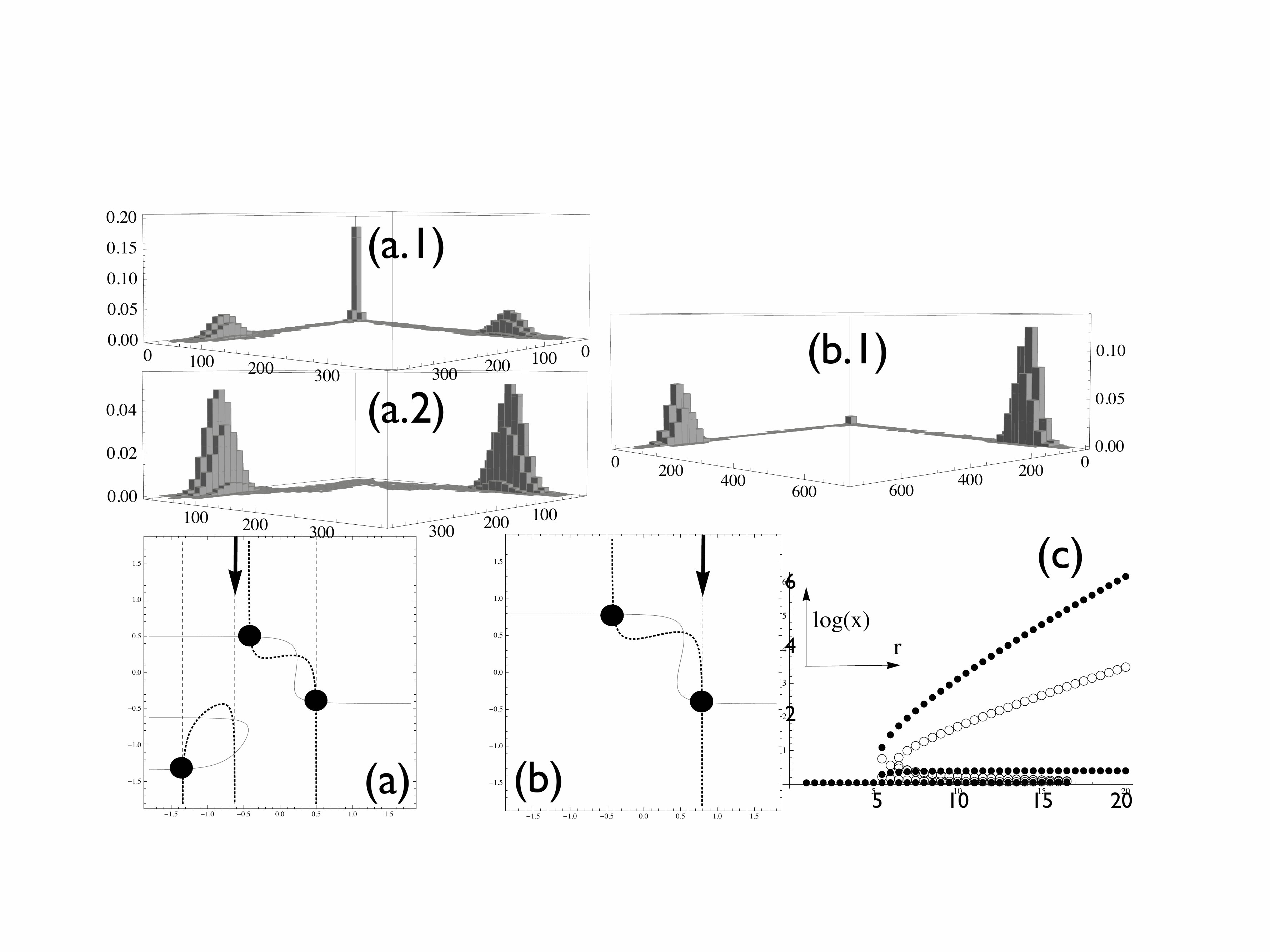}
 \end{center}
 \caption{{\bf The (heterozygous) context dependent dual-activator switches between three different expression states.}  The $x_1$-$x_2$ planes in this figure are in logarithmic scale.  For the case $r_{11}=r_{22}=r_1$, $r_{21}=r_{12}=r_2$, we obtain multiple steady states.  Here we illustrate the transition
    between three and two steady states as the ratio $r=r_1/r_2$ changes
    from a value of 10 in (a.1) and (a.2) to 20 in (b.1).  All other parameter values are as in \ref{parameters},  with $c=\Delta=1$, and the activation strengths are multiples, $r$ of $r_0$. The two histograms in (a.1) and (a.2) correspond to different initial conditions on either side of the arrow drawn in the phase-plane (a) for small noise (large $\Omega$). The vertical lines in (a) and (b) correspond to the case of the single gene switch.  The outward point double-arrow in (a) shows the threshold for the single gene case (which is close to that of the duplicated gene as well), while the inward pointing single arrow locates the only (stable) fixed point for large $r$ in the single gene case.  On the right are shown the stable and unstable fixed points of $x_1,x_2$ (ordinate) for different values of $r$ (abscissa).  Because of symmetry $x_1$ and $x_2$ share the same steady state values, but not concurrently. }
     \label{fig:hetswitch}
\end{figure}

    The fixed points are now given by intersections of cubics obtained by setting $({d}/{dt}){{x_{1,2}}}=0$ in eq. )\ref{eq:ODEmodel}), where we ignore the downstream sector.   We find multi-stable solutions in this case -- a tri-stable state with $(x_1^*,x_2^*)$ levels being (low, low), (low, high) and (high, low) which undergoes a change via a pitchfork bifurcation upon increasing $r$ to a bi-stable state where the low expression state is lost and only the
mutually exclusive expression states (low, high) and (high, low) for $(x_1^*,x_2^*)$ remain. Figure \ref{fig:hetswitch} illustrates this.    

%

\begin{figure}[ht]
  \begin{center}
\subfloat[]{\includegraphics[width=0.45\textwidth]{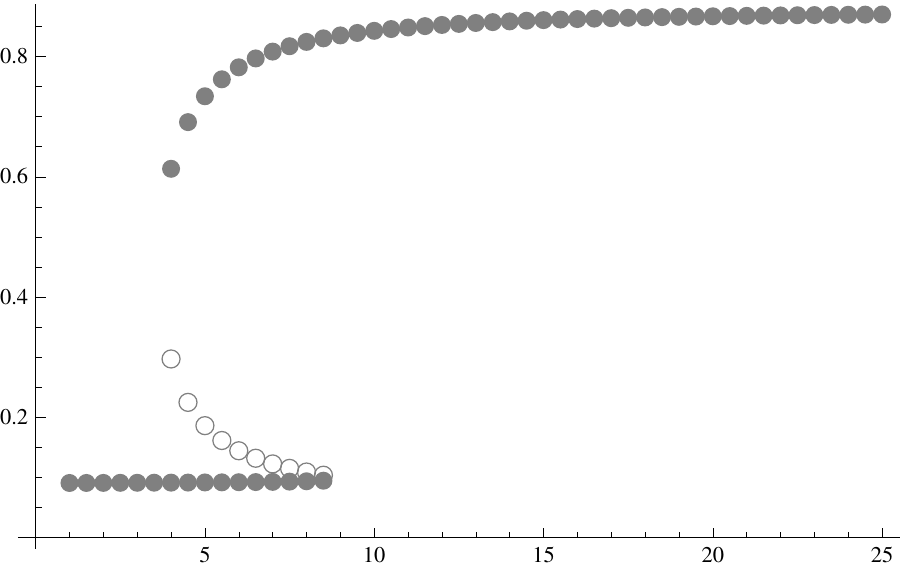}}
\subfloat[]{\includegraphics[width=0.45\textwidth]{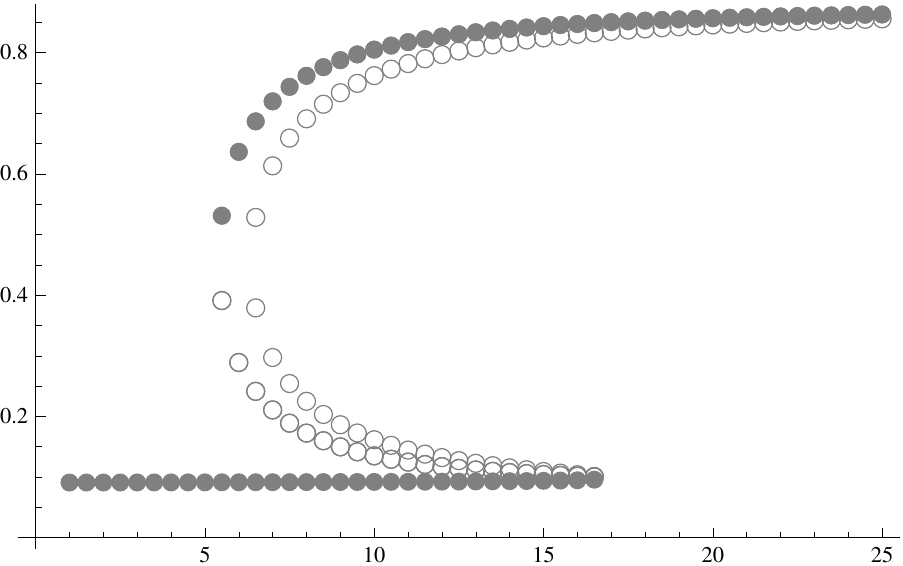}}\\
\subfloat[]{\includegraphics[width=0.45\textwidth]{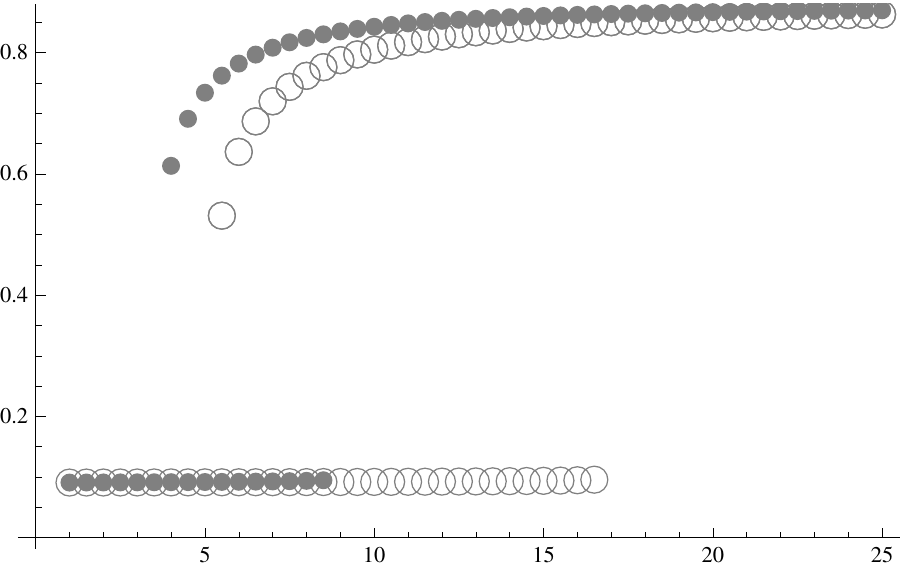}}
\subfloat[]{\includegraphics[width=0.45\textwidth]{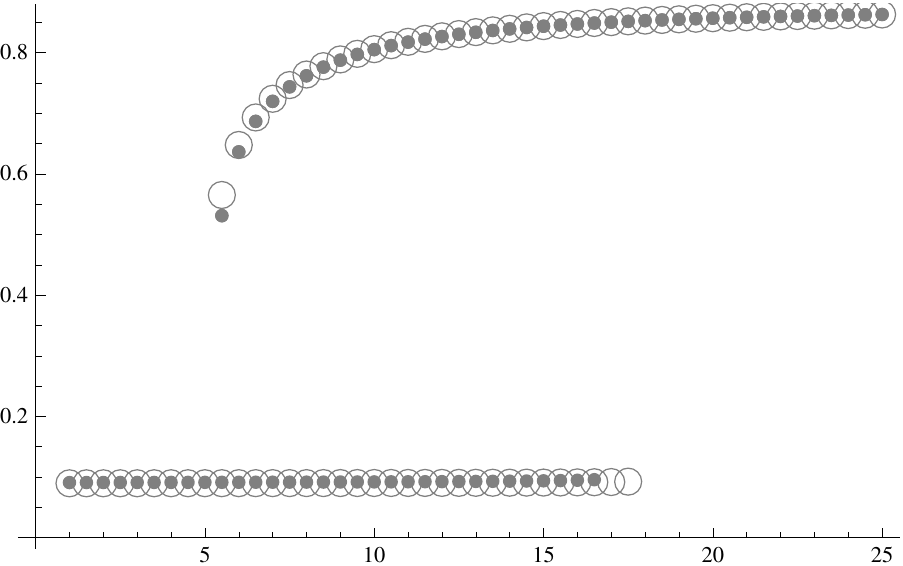}}
\end{center}
 \caption{{\bf (Expression levels vs $r$).  The decision switch is indistinguishable from the single gene switch in downstream effects.} In each of these figures, the abscissa indicates the value of $r$ in the two-gene subnetwork.  In (a) we have the case of unchanged duplicated autoregulatory gene $r_{ij}=r$, while in (b) we have the case $r_{11}=r_{22}=r$ while $r_{12}=10r_{0}=r_{21}$ where the basal rates are taken to be the same $r_{i0}=r_0$.  The open circles in the top two figures (a) and (b) are the values of the target gene $z$ corresponding to the unstable fixed points.  The stable fixed point values are in filled circles.  The stable values of $z$ in (a) and (b) are gathered together in (c) for comparison, with the open circles correspond to values in (b), filled circles to those in (a).  Finally, in (d) we compare the stable values from the context-dependent case (b) to those from the target expression of the pre-duplicated system.  Note that if we use the expression of the target gene as a readout of the effect of gene duplication, we find no change (see (d)) prior to further divergence.  For the decision switch, (c) shows that the only qualitative change lies in the threshold for switching and hysteresis.   }   
  \label{fig:compareswitches}
\end{figure}

We have thus far studied the effects of complementary mutations following duplication that affect $r_{ij}$ by affecting protein-protein interactions at the enhancers and thus $\varepsilon_{hA_j p}$ in eq. (\ref{eq:deftherm}), where helper proteins indexed by $h$ provide transcriptional response specificity in this modelling framework.  $h$ is a place-holder for the \emph{cis}-regulatory context of gene expression.  In order to implement the complementary loss-of-function mutations  
of \cite{force-genetics1999}, we have the helper protein enhance the recruitment of transcriptional machinery 
in complementary ways.  Starting from $\vect{r}_i=(r_{i0},r_{i1},r_{i2})$ with equal $r_{i1}$ and $r_{i2}$, we end up with a situation where transcription from locus $1$ is greater for the coding region $1$ (say) than the coding sequence $2$ ($r_{11}>r_{12}$) and similarly,  $r_{22}>r_{21}$ (details in \ref{context2}).   We compare the two cases of  progressive (homozygous) and decision (heterozygous) switches \cite{guantes-poyatos-switch} in Figure \ref{fig:compareswitches}. 

We use the expression level of a downstream gene as an indicator variable (abscissa in Figure \ref{fig:compareswitches}) to locate the influence of the control parameter $r$, the ratio of activation strengths of the two transcription factors (ordinate in Figure \ref{fig:compareswitches}).   The filled circles correspond to stable fixed points of the expression of a downstream gene activated by the pair of duplicated activators and the open circles are those that correspond to the unstable fixed points identified in Figures \ref{fig:homswitch}, \ref{fig:hetswitch}.  In Figure \ref{fig:compareswitches} (c) we show that the key difference in the two cases (now the open circles correspond to the closed circles, or stable points in (b), and the filled circles are the stable steady states in the duplicated-no-divergence case of (a)) lies in the extent of hysteresis that the system affords.  The doubled autoregulator shifts the threshold for entry into the high-expression state lower compared to the pre-duplicated gene.  The asymmetrically expressed decision switch case is compared to the pre-duplicated situation in 
Figure \ref{fig:compareswitches} (d), with barely any difference in the expression levels in the two cases.

The doubling of the self-activating gene comes with an obvious corollary in that the system retains one copy of the switch upon deletion of one of the two copies of the gene.  Thus, even in the case that a gene is in a low-expression state as in Figure \ref{fig:hetswitch} (b), the deletion of its high expression partner induces its upregulation.  Such a back-up feature points to the plausibility of invoking robustness in differentiating between the systems yielding the two expression levels in Figure \ref{fig:compareswitches} (d), which are otherwise indistinguishable.  This stable dynamical state is of course, unstable from an evolutionary perspective unless the conditions that lead to the potential loss of a paralog are heritable as well. 

An hypothesised original autoregulatory developmental gene has been used to replace two copies of the \emph{hox1a,b} paralogs in a mouse with consequent minor alterations to normal development of its forebrain\cite{hox1reversal}.   However, in the wild-type mouse, the prominent phenotypes that develop upon knock-out of a paralog means that this back-up facility that is a consequence of duplication has been lost, via the loss of the \emph{hoxa1} autoregulatory site, and acquisition of further specialised roles for the proteins.  This dynamically stable state of the duplicated autoregulator is rendered evolutionary stable by the divergence of the pleiotropic roles taken on by the duplicates, which are split and stabilised by the model of subfunctionalization \cite{force-genetics1999,stoltzfus99}.  An example of a developmental system where genetic buffering upon deletion of a paralog is observed is the \emph{myoD} and \emph{myf5} pair involved in myogenesis where deletion of  the pair induces the upregulation by its partner\cite{wang-redundancy96}.  However, this appears not to be a cell-autonomous property: results from clonal lines \emph{in vitro} do not show the same effects as in the organism, and further studies \cite{haldar08} indicate lineage-specific divergence of gene function, with buffering occurring at an inter-cellular stage.

\section{Emergence of oscillations due to changes to activation site}

The standard model of gene activation \cite{ptashne-gann-book} involves recruitment of Pol II and other transcriptional proteins by protein-protein interactions with the activation domain of the transcriptional activator.  Increasing the activity of the (acidic) activation domain is also accompanied by an increase in the protein degradation rate \cite{wang-ptashne-unstable2010}, a correlate that is key to the results below.  Mutations to the activation domain to either copy of the duplicated gene/allele can thus introduce a change in the time scales of the dynamics of the two proteins.  If we view the enhancer region of the gene as the source from which proteins are produced, duplication followed by alteration of the activation region introduces competition for this source.  Such a competitive framework arises in ecological theory where one species can take over a food source or habitat at the expense of another -- a property called competitive exclusion.  However, it has been shown \cite{armstrong-exclusion80} that it is possible for two competing species of predators to subsist on a single species of prey (a food source) in an oscillatory mode.   This analogy extends to our model as well. 

For the case where the divergence between copies is modular both enhancers are taken to be copies of each other with $t_{ij}=1$ and the occupancy of the promoters are the same, $\varphi_1=\varphi_2=\varphi$.  In detail, modular activation implies $r_{ij}=r_{\circ j}$, independent of genomic locus $i$.  Thus, we take $$\vect{r}_i=(r_{i0},r_{i1},r_{i2})=(r_{\circ 0},r_{\circ 1},r_{\circ 2})=:(r_{0},r_{1},r_{2})$$ 
in this section.  
 The fixed points $(x_1^*,
x_2^*)$ are easily obtained by setting $(d/dt)({x_1}, {x_2})=(0,0)$ in eq. (\ref{eq:ODEmodel}) to
find $(x_1^*/x_2^*)=(c_1\Delta_2/c_2\Delta_1)$, so that 
\begin{equation}
c_1\varphi(x_1^*, (c_2\Delta_1/c_1\Delta_2)x_1^*)-\Delta_1 x_1^* = 0
\end{equation}
is a cubic equation with either 1 or 3 real solutions.  Since we can
set a rescaled time variable, there are principally 3 parameters that
determine the different dynamical outcomes of this model.  We shall
adjust the ratio of the decay rates $\Delta=\Delta_1/\Delta_2$ to capture the different time-scales for the two copies, the ratio $c=c_1/c_2$ which includes the number of binding sites and can be used to model diploid versions of the duplicated haplotype, and $r=r_1/r_2$ the relative affinities for the activators to the
transcriptional machinery.  

\subsection{Oscillations by local and global bifurcations}

The stability of a fixed point is usually investigated by the
behaviour of the vector-field in its vicinity, \emph{i.e.}, by
observing how the system responds to a perturbation about that point.  Using the factorisation eq. (\ref{eq:factored-det}) of the Jacobian of the full system eq. (\ref{eq:ODEmodel}) we focus our interest on the two-activator subnetwork as being the generator of novelty rather than the dosage dependence downstream.  In particular, we seek out the conditions for a Hopf bifurcation to find oscillatory solutions to the equations (see \ref{elimination}), using  Sylvester resultants \cite{cox-iva}. 

 \begin{figure}[ht]
  \begin{center}
\subfloat[]{\includegraphics[width=0.5\textwidth, angle=0]{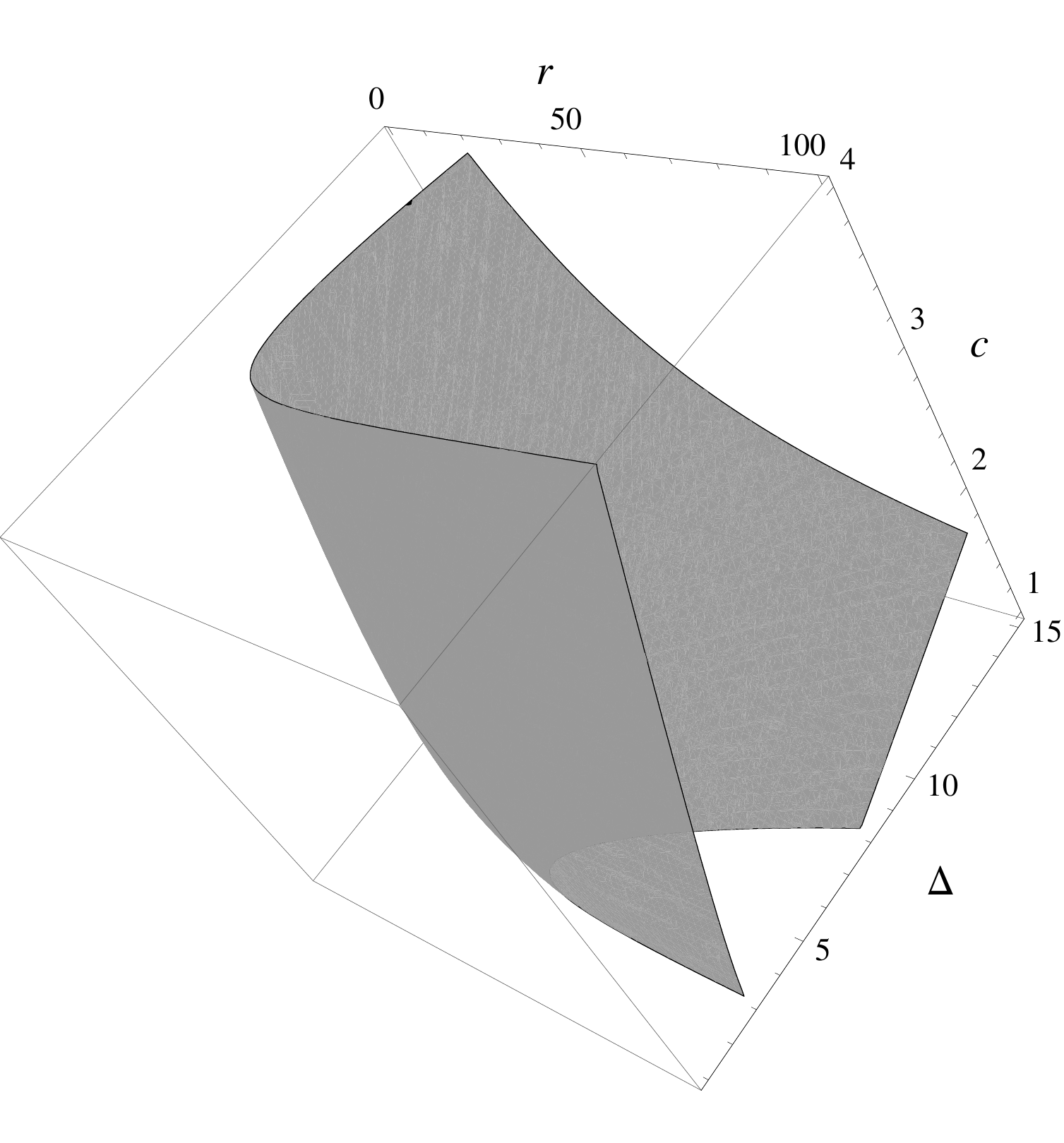}}
\subfloat[]{\includegraphics[width=0.25\textwidth, angle=0]{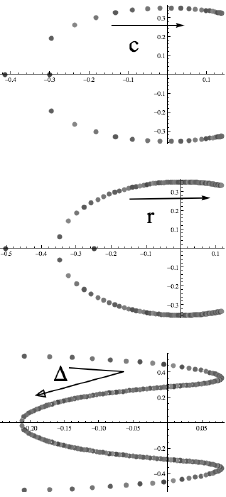}}
 \end{center}
 \caption{{\bf Parameter space where Hopf bifurcation occurs.}   As explained in the text, eliminating $x_2$ from $f_1$ and $f_2$ for a set of parameter values $c_2=50$, $r_0=10^{-3}$ and $r_2=10^{-2}$ yields a surface in $(r,\Delta,c)$ 3-dimensional parameter space depicted in (a).  The rest of the parameters are as in \ref{parameters}. In (b) we show the complex plane of the eigenvalues with two of the three parameters fixed, and the increase of the third through the bifurcation point(s).  This occurs as $c,r$ increases, and the non-monotonic increase of the positive real part of the eigenvalue as $\Delta$ increases is also clearly seen at the bottom of (a).  } 
\label{fig:hopfsurface}
 \end{figure}

\begin{figure}[htbp]
   \centering
   \includegraphics[width=.6\textwidth]{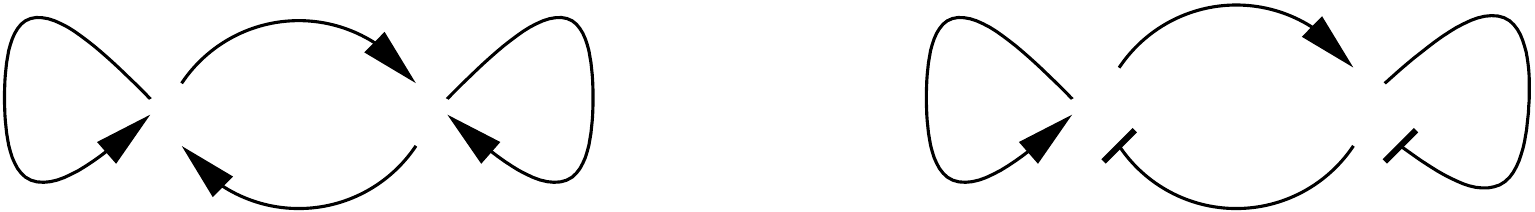} 
   \caption{The two topologies obtained by solving for the Hopf bifurcation condition.  The one on the left is the subnetwork we address in this paper.}
   \label{fig:topologies}
\end{figure}

There, we find two conditions for a Hopf bifurcation, by factoring a polynomial constraint into two pieces.  First, 
\[
\frac{c}{\Delta}=\sqrt{\displaystyle-\frac{r_2-r_0}{r_1-r_0}}
\]
which implies that one of the two genes recruits Pol II more efficiently than the basal rate while the other's activation rate is less than that of basal transcription -- \emph{i.e.}, one is an activator, the other a repressor.  This topology was proposed in  \cite{hasty-prl02} and implemented in \cite{stricker08} and shown on the right in Figure \ref{fig:topologies}.   
We do not deal with this case in this paper.  

The second factor in the polynomial constraint obtained by the Sylvester resultant in \ref{elimination} gives rise to the model that we work with in this paper, shown on the left in Figure \ref{fig:topologies}.  This factor gives a surface in the 3-dimensional space 
determined by ($(r_2/r_1)$, $(c_2/c_1)$, $(\Delta_1/\Delta_2)$) as shown in Figure \ref{fig:hopfsurface} 
for chosen values of $\Delta_2$, $c_2$ and $r_0$.  This demonstrates that a two-activator system can sustain stable oscillations and illustrates how topology is an insufficient predictor of network function.  As 
suggested by the observation of dual regulation by $x_2$ as a function
of $x_1$ levels, low-expression states participate in a positive
feedback loop that maintain a stable steady state, whereas for high
$x_1$ levels $x_2$ behaves like a repressor and the oscillatory
behaviour of positive and negative loops \cite{hasty-prl02} is
observed.

\begin{figure}[ht]
 \begin{center}
\subfloat[]{\includegraphics[width=0.45\textwidth, angle=0]{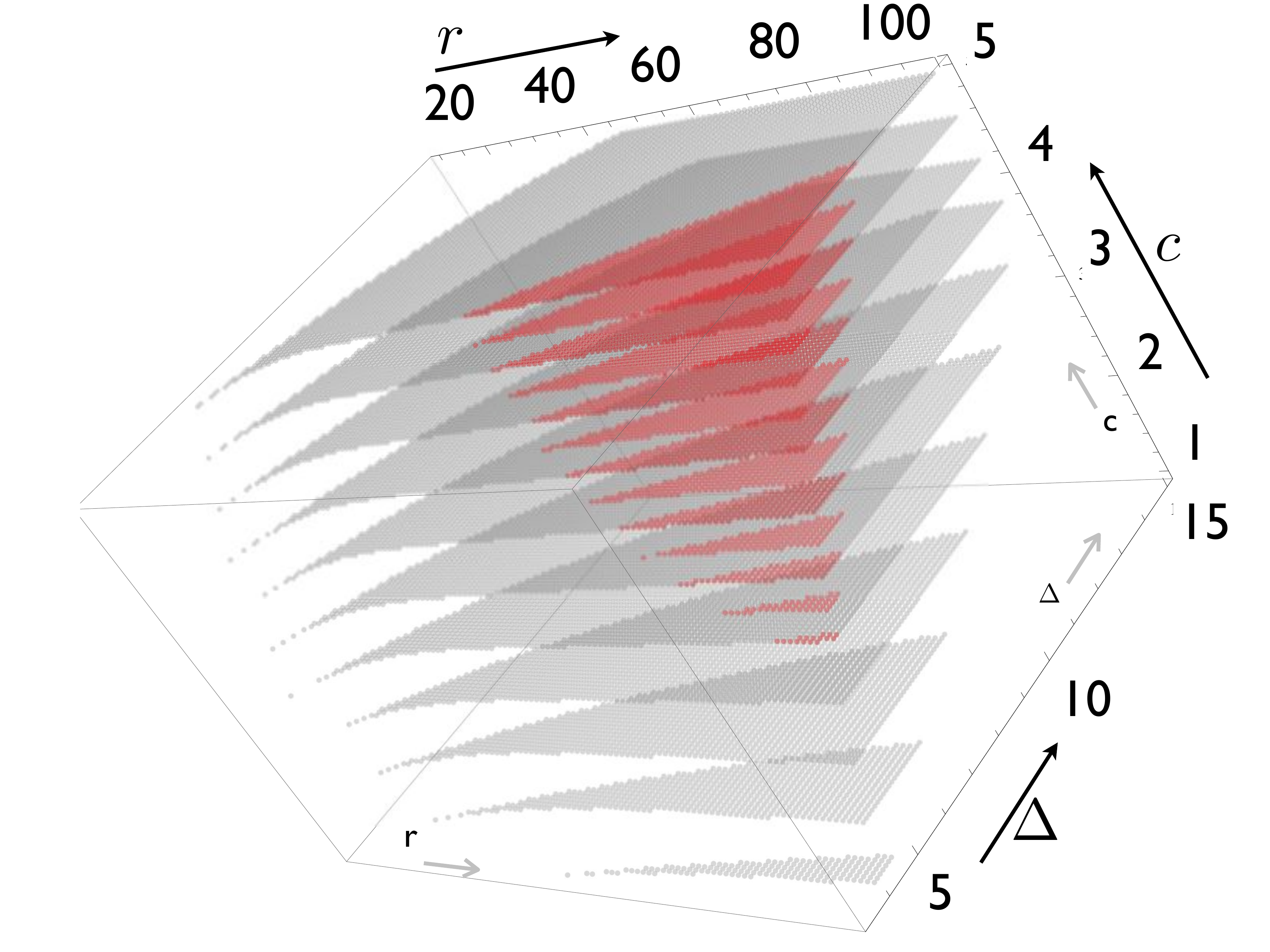}}
\subfloat[]{\includegraphics[width=0.5\textwidth, angle=0]{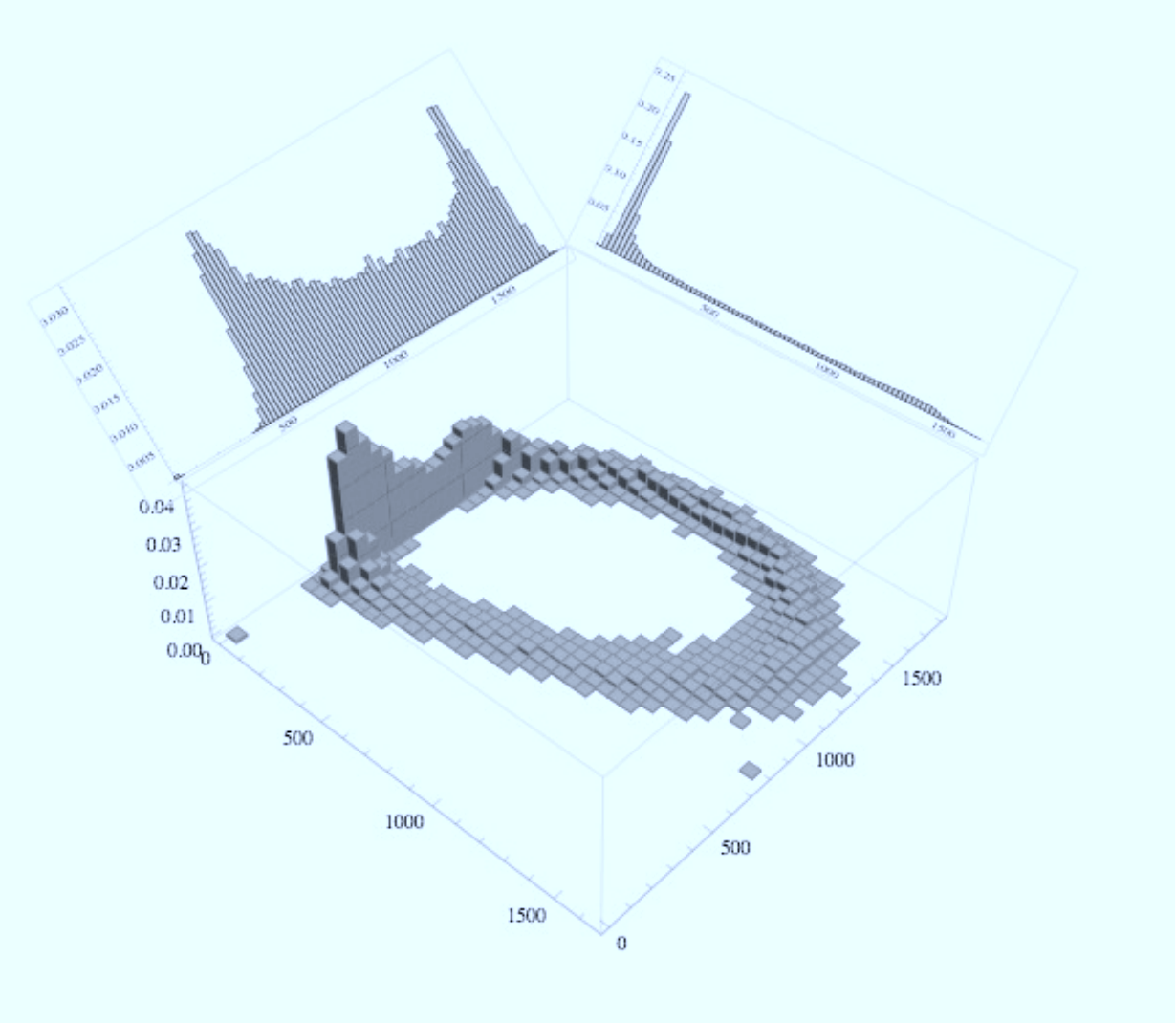}}
\\
\subfloat[]{\includegraphics[width=0.4\textwidth, angle=0]{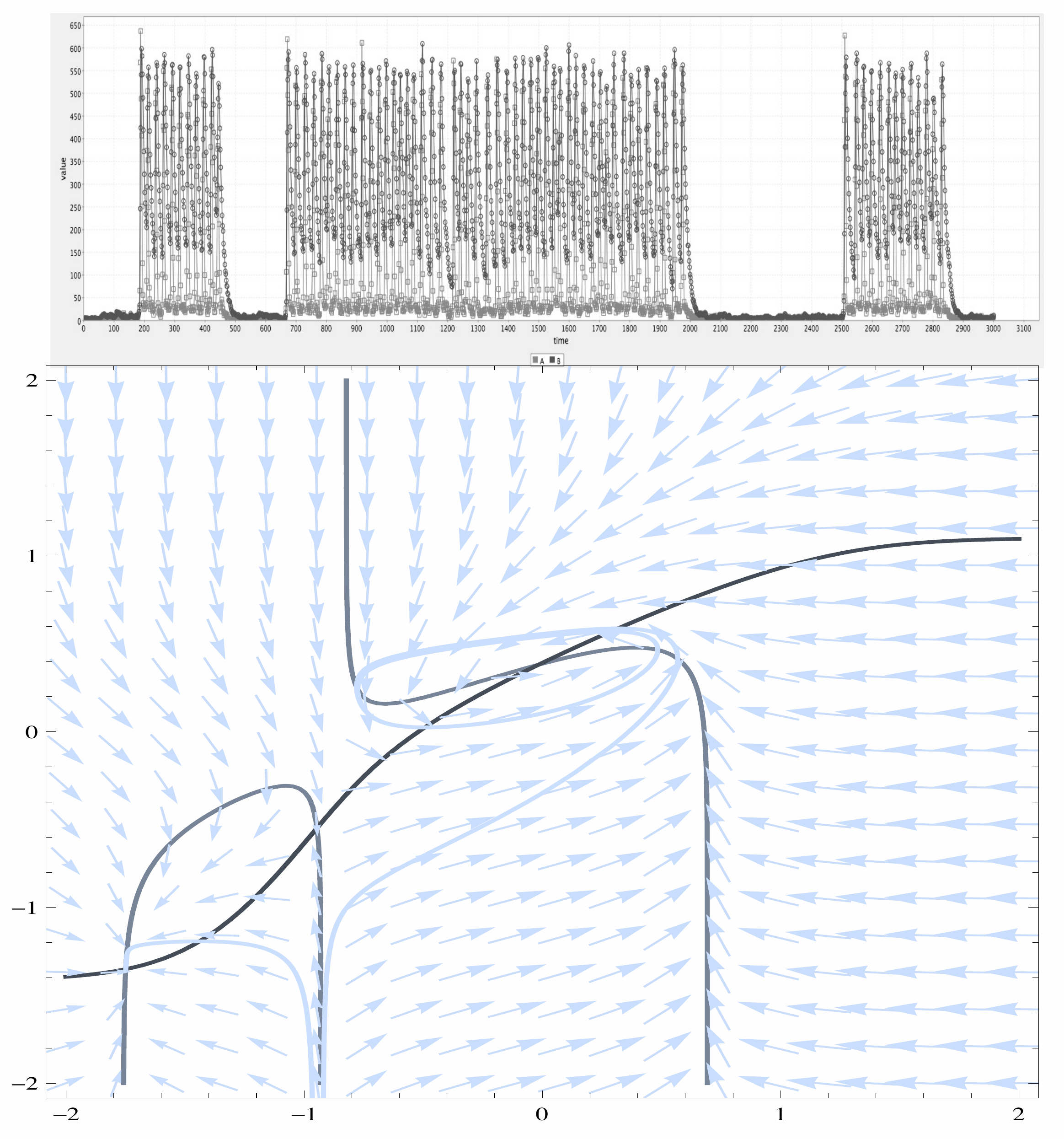}}
\subfloat[]{\includegraphics[width=0.4\textwidth, angle=0]{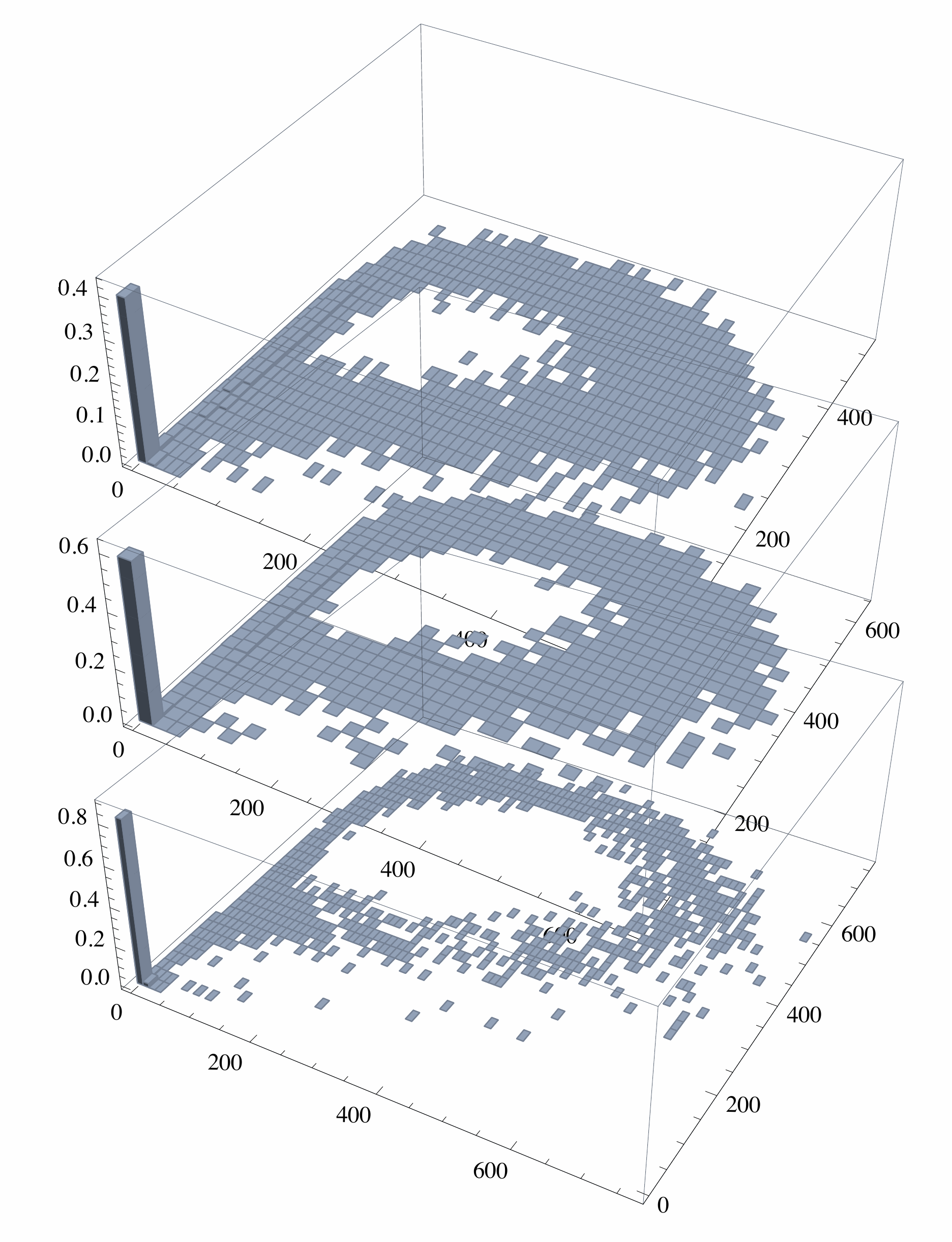}}\end{center}
 \caption{{\bf Bursts of oscillations are controlled by the amount of noise in expression.}  The $x_1$-$x_2$ planes in this figure are in logarithmic
   scale. (a): Oscillatory regimes in parameter space -- red (dark online) region has a stable limit cycle; the light (grey online) has a coexisting steady state as well. 
 (b):Histograms accumulated from a simulation of the stochastic
   kinetics in the red oscillatory phase with ($x_1, x_2, P(x_1, x_2)$) as the right-handed coordinate axes.
  Also shown are histograms for $x_1$ (unimodal) and $x_2$
   (bimodal). 
   (c):    Phase plane with $x_1-x_2$ nullclines and the trajectories of two
   different initial points, chosen to be on either side of
   the separatrix determined by the saddle-node (in the middle of the three intersections).  
   (d):   Histograms of trajectories in the light grey region of (a) for different noise strengths
    (which decreases from top to bottom) showing the redistribution of probability mass around
   the limit cycle away from low-expression peak.   }  
\label{fig:osc-all}
 \end{figure}

The subset of the parameter ranges for which the system oscillates in Figure \ref{fig:osc-all} (a) where the light grey and dark grey (red online) regions show different kinds of stable dynamics.  The dark (red online) region displays a region where there is only one fixed point with stable oscillations as shown in Figure \ref{fig:osc-all} (b) even for a stochastic version of the model (see below).  The light grey region is where a stable limit cycle and a steady state coexist. 
To give an indication of the biological plausibility of such a mechanism, the biophysical parameter of interest is the ratio of binding rates to the transcriptional machinery, $r=r_1/r_2$.  To generate oscillations, a value around $r=50$ is sufficient for small values of $c$.  This translates into a free-energy of binding differential of $\Delta\Delta G\sim 2.3$ kcal/mol.   

\begin{figure}[htbp]
   \centering
   \includegraphics[width=0.23\textwidth, angle=90]{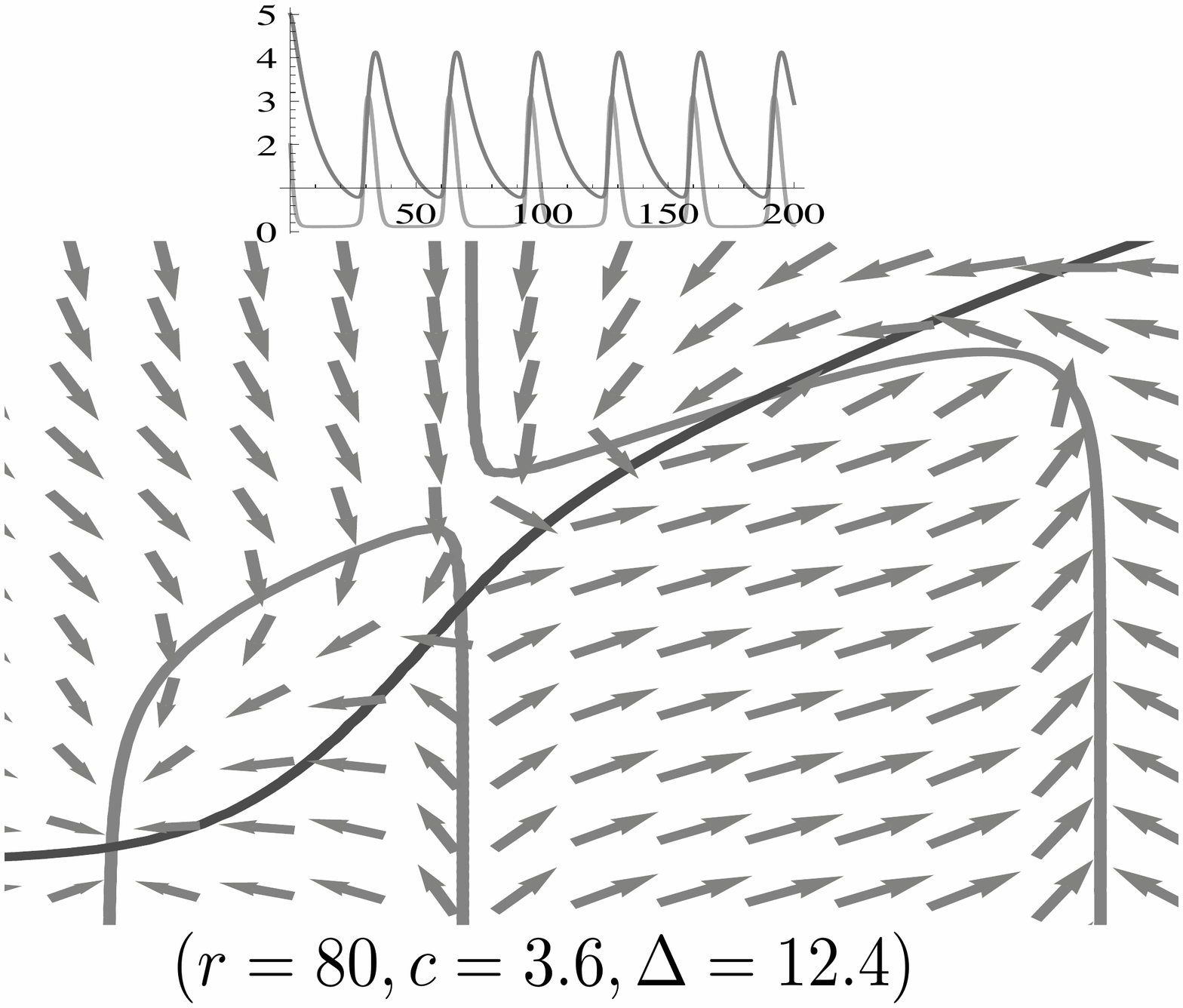} 
   \includegraphics[width=0.23\textwidth, angle=90]{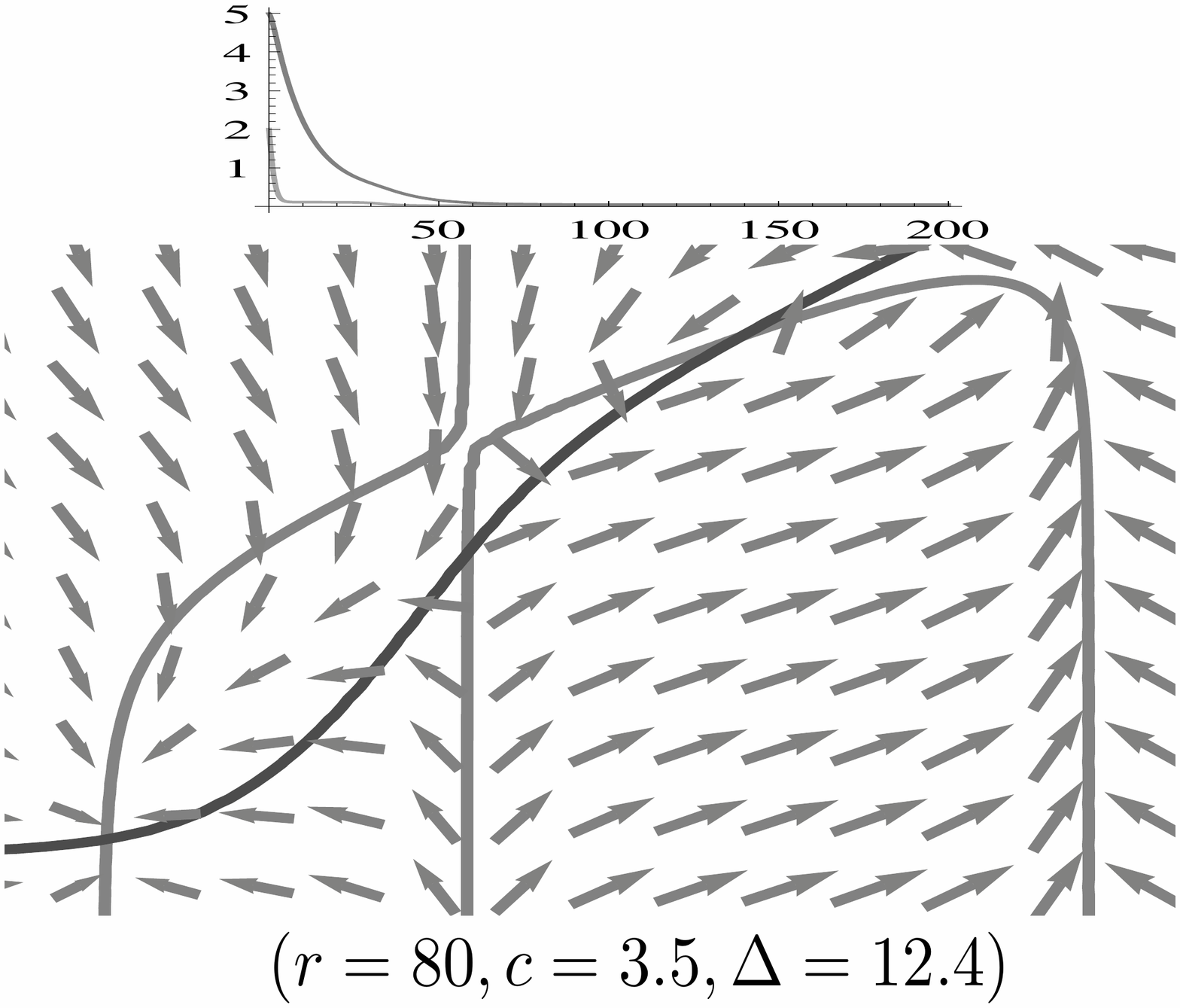} 
   \includegraphics[width=0.23\textwidth, angle=90]{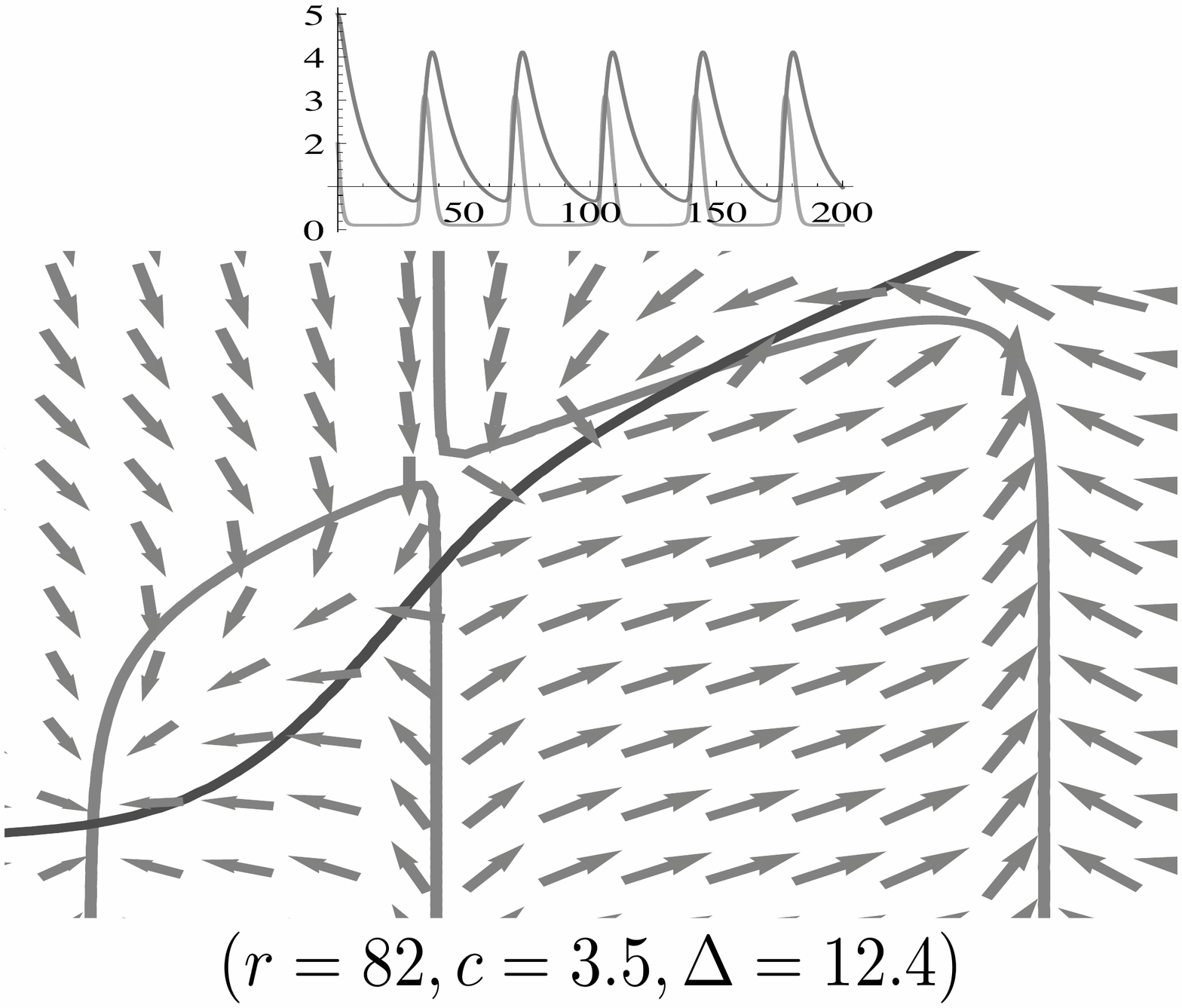} 
   \caption{{\bf Global bifurcations} leading to oscillations are exemplified in this figure, where an excitatory trajectory can be trapped into an oscillatory state by realignment of nullclines upon parameter changes.  This can happen for both, changes of $r$ and of $c$: the middle panel has parameter values  $\theta^*=(r=80, c=3.5, \Delta=12.4)$, whereas the left panel has $\theta+\delta c$ and the right panel has $\theta+\delta r$, with $\delta c = 0.1$ and $\delta r = 2$.  Either change modifies the $x_1$ nullcline from $)(\rightarrow\genfrac{}{}{0pt}{}{\cup}{\cap}$. Time series plots for each case is shown above the phase planes.}
   \label{fig:global}
\end{figure}

We also point out that there are global bifurcations in this model that lead to the emergence of oscillations upon parameter changes, as shown in Figure \ref{fig:global}.

\subsection{Coexistence of one stable equilibrium and one oscillatory state}

The presence of a saddle node separating a stable steady state and an
unstable fixed point with complex eigenvalues enables the system to be
at either of these two states.  Below the threshold set by the saddle,
the system settles into the stable steady state; above it, the system
oscillates.   A saddle-node bifurcation arises when the stable fixed point and 
the saddle merge \cite{strogatzbook} (the light-dark (grey-red online) boundary in Figure 
\ref{fig:osc-all}(a)), leaving the system to exist only 
its oscillatory state.   

%
%
%

The coexistence of an equilibrium state and a stable limit cycle
(the light (grey) volume shown in Figure  \ref{fig:osc-all})
enables oscillations to be annihilated by the suitably chosen
perturbation \cite{winfree} as well as for rhythms to be switched back
on, but with a reset phase, by a threshold-crossing perturbation
around the stable fixed point.  Such a mechanism underlies a temperature compensation scheme in circadian clocks in flies \cite{hong-tempcomp-pnas07}, suggesting plausibility arguments for fitness parameters that favour such a phenotype.

\subsection{Noise-driven oscillations and bursts}

The coexistence of stable equilibrium and stable limit cycle dynamics
in the deterministic model suggests that a stochastic version of the
model, incorporating intrinsic noise of biochemical reaction steps
that constitute the network, will show switching behaviour between
these phenotypes.  The stochastic kinetic model (which is a simplified
reduction of the detailed set of molecular reactions presented in \ref{kinetic} ) contains a ``volume parameter'' $\Omega$
which enables the translation from (nano-)molar concentrations used in
deterministic chemical kinetics models implemented as ODEs and numbers
of molecules that are present in cellular volumes.  We have generated
sample paths using Gillespie's algorithm \cite{gillespie76} for different values of
$\Omega$ to observe the extent of noise in the trajectories as well as
the frequency of switching between these states.  We illustrate the
occurrence of noise driven transitions in Figure  \ref{fig:osc-all}(c,d).  By modulating
the threshold by an external input it would also be possible to
generate frequency modulated bursts of oscillations, an additional
mode of dynamical processing made available to the cell via such a
double-activator subnetwork.

\subsection{Population levels are at steady state even when cells oscillate}

The bursty oscillatory character of the system reflects noise-driven switching between a stable steady state and a high mean-expression oscillatory state.  The random nature of the switching resets the phase and the lack of coherence in an unsynchronised population between individual oscillators results in very low amplitude oscillations.  This averaged behaviour is not significantly different from a steady state population response.  Furthermore, univariate histograms for the system that displays stable oscillations or oscillatory bursts display bimodal distributions of proteins reminiscent of switches.   Thus, it is plausible that a population of cells could  present a responsive interface to environmental inputs, such as driving metabolic processing of nutrients in a manner that is neutral, and hides the novel time-dependence in individual cells.    This might enable anticipatory responses to periodic environmental cues by processes further downstream to develop and manifest themselves, or indeed be stabilised via duplication events in diploid species (see below).  The smaller parameter range shown in Figure \ref{fig:osc-all} when the steady state is lost and only the oscillatory state remains indicates that it is possible for the system to mask the potentially harmful acquisition of oscillatory instability and commit to oscillations under the future favourable environmental conditions for its subsequent selection.

\section{Evolution of duplicates for the oscillating phenotype}

The outcome of introducing a duplicate mutant will be to either successfully invade the resident population of singleton alleles or die out.  In this paper we do not explicitly model the effects of drift, but we note that in small populations, selection coefficients much smaller than $(1/N_e)$, where $N_e$ is the effective population size will get swamped by drift.  Moreover, while there were two scenarios explored in the dynamics of regulation -- a switch-like or an oscillatory phenotype -- we shall only explore the evolutionary consequences of oscillatory dynamics and its role in the spread of the duplicate gene in this paper.  Recall, the fitness coefficients in (\ref{tab:tablefitness}) are organised to explore the consequences of increased fitness arising from unequal activation strengths of the two alleles of a self-activating gene upon the potential for retaining gene duplicates.  In the singleton case, homozygotes $a_1/b_1$ or $a_2/b_2$ (note: $a_2$ and $b_2$ are the same allele) have $r_1=r_2$ and $c_1=c_2$, and hence the phenotype is non-oscillating.  The allocation of fitness contributions $(1-s)$ and $(1-t)$ imply that there is positive selection for the heterozygote singleton.  While this may arise for a variety of factors, we have identified the onset of oscillations in both deterministic and stochastic versions of the regulatory network model as the main qualitative difference.  Hence, we investigate the potential for the duplicate to increase in frequency under conditions that do not require additional fitness advantages to doubling $c:=c_1/c_2$ (\ref{eq:cidef}), setting $d=0$ in (\ref{tab:tablefitness}).  (One can see in Figure \ref{fig:osc-all} that increasing $c=c_1/c_2$ pushes the system deeper into the oscillatory region, and indeed, can drive a Hopf bifurcation, as seen in Figure \ref{fig:hopfsurface}.)  One question we investigate here is whether halving $c$ in (\ref{eq:cidef}) in the genotypes $a_1b_0/a_2b_2$, $a_2b_0/a_1b_2$ and $a_2b_0/a_2b_1$, which has the phenotypic effect of reducing the propensity for oscillations, affects the potential for the duplicate gene to spread. For this, we set $u=t+u'$, where $u'=0$ is the condition that it matches the fitness of homozygote singleton $a_2b_2$ genotype, and do a perturbative analysis around $u=t$.

Since both $2$-by-$2$ recombination matrices in (\ref{eq:admatrices}) are positive, their largest eigenvalue has a unique corresponding eigenvector (called the Perron eigenvector) which has all entries of the same sign (taken positive), by the Perron-Frobenius theorem.  If the corresponding eigenvalue is greater than $1$, the linear combination of haplotypes defined by the entries of the Perron eigenvector increases in frequency due to the process of recombination with the existing singleton haplotypes in equilibrium.

 The $\mathbf{R_{11\times 21}}$ recombination map has a largest eigenpair $(\lambda,\vect{v})$ to first order in $(t-u)$ and $d$:
\begin{equation}
\label{eq:eigad}
\begin{array}{rcl}
\lambda-1&=&\displaystyle\frac{s^2}{\hat{W}(s+t)^2}\left((t-u)+d\left(\frac{t}{s}\right)+\cdots\right) \\ 
\vect{v}&=&\left(\begin{array}{c}\displaystyle \frac{t}{s}-(t-u)\frac{t}{\rho(s+t)}-d\frac{t(t-s)}{\rho s(t+s)}+\cdots\\1\end{array}\right).
\end{array}
\end{equation}
This shows the relative evolutionary consequences of dosage \emph{balance} between the alternative alleles $a_1$ and $a_2$ by noting the unequal contributions of $[\phi_1]=2[\phi_2]$ and $2 [\phi_1]=[\phi_2]$ to fitness.  If the phenotype of oscillations is favoured, we have $[\phi_1]=2[\phi_2]$, whereby $W_{10;21}=W_{10;12}=W_{20;11}=1+d$ with $d\geq 0$; $d>0$ indicates the positive benefits of dosage imbalance and thereby increasing the amplitude of oscillations (data on amplitudes not shown).  If $2 [\phi_1]=[\phi_2]$, $c\rightarrow (c/2)$ upon duplication, which reduces the likelihood of oscillations (see Figure \ref{fig:hopfsurface}).   If the value of $c$ were sufficiently high preceding duplication (due to the various factors in eq. (\ref{eq:cidef})), this might yet permit the system to retain an oscillatory phenotype. leading to the condition $t>u$.  Hence, eq. (\ref{eq:eigad}) shows that no additional benefit has to accrue to the oscillatory phenotype ($d=0$) for the duplicate to propagate if pre-duplication allelic divergence is selected for\cite{spofford1969,OttoYong2002,proulx2006}. The high-amplitude oscillatory state where $[\phi_1]=2[\phi_2]$ may indeed be weakly selected against ($d<0$) and yet enable to the spread of the duplicate.  Consequently, the proportion of the duplicated allele $a_1b_1$ is reduced compared to $a_2b_1$ as evident from the Perron eigenvector.  This reduction is larger for tightly linked genes (small $\rho$, $0\leq \rho\leq (1/2)$).  This suggests different consequences for duplicated genes whether they arise by tandem duplication or by retrotransposition.

\section{Discussion}

In this paper, we find that duplication of an autoregulatory gene can lead to competition for common genomic binding sites.  Assuming a cooperative mode of regulation described via, but not requiring, a dimeric mechanism,
we find that a dual regulatory behaviour -- context dependent activation or repression -- ensues.  Feedback amplification of dual regulation gives rise to switching and oscillatory behaviour.   While a minimal model like the one we have chosen involves an autoregulatory loop, we anticipate similar qualitative changes to emerge in the more general context of duplication of the node closing a larger feedback loop.   The model displays  mutually exclusive expression levels under appropriate contextual cues, analogous to those specified by the subfunctionalization model for retention of duplicates.  When the 2-node double-activator subnetwork is viewed as alleles in a diploid species, allelic divergence of expression emerges in this case.  We note, in this context, that partitioning of expression of different alleles occurs in $F_1$ hybrid cotton plants for the alcohol dehydrogenase \emph{Adh} gene \cite{AdamsF1Silence2005}.  Moreover, we also find that the ensuing circuit is capable of back-up of expression of a deleted gene by its paralog {kafri-pnas06}.  This is linked to developmental systems where such a phenomenon has been observed.  The evolutionary instability of this genetic buffering has not been modelled, however,  as we have not included the effects of mutational loss in our evolutionary analysis.  

We have constructed fitness functions that favours the heterozygote, implying that coupling to periodicity in the environment enhances the selective advantage of an organism bearing such a duplicate.  In an \emph{in silico} evolution experiment it has been shown that in a periodic environment increased metabolic flux is obtained via glycolytic oscillations when compared to a non-oscillating response \cite{tsuchiya-ross-pnas2003}.  Further, it has been experimentally realised that hexose transporter genes regulating influx of sugar is crucial for the establishment of yeast glycolytic oscillations \cite{reijenga-biophysJ2001}, and yeast evolves to outcompete ancestral colonies in nutrient-limited conditions by duplicating hexose transporter genes including \emph{HXT6/HXT7} \cite{brownhxt98}.  The coupling of oscillatory phenotype to environmental cues has been most often discussed in the context of circadian rhythms \cite{mcknight-conjecture-annurevbiochem02,DoddPlantCirc2005}.  The results of \cite{hybrid-circadian2009} show that 
increased biomass yields of hybrids \cite{heterosis2010} -- $F_1$ (allopolyploid) crosses of two \emph{Arabidopsis} species -- was correlated with increases in amplitude of circadian rhythms of clock and clock-controlled genes.  While $F_1$ hybrids are often evolutionarily unstable, our simple evolutionary model shows the spread of a duplicate in such a system with overdominance.  Circadian rhythm generating mechanisms have been much studied, and it has been shown in mammalian systems that competitive binding to common conserved \emph{cis}-regulatory regions (such as the E/E'-box, the RRE and the D-box in mammals \cite{ukai-synthesis-circadian08}) drives circadian rhythms, and that its removal reduces the amplitude of oscillations.  While this appears to be consonant with our findings, it should be pointed out competitive binding is the sole mechanism in our model, unlike in the circadian clock.  The competitive mechanism for the generation of oscillations relies on the increased efficacy of activation of one of the activators feeds greater proteolytic rates \cite{wang-ptashne-unstable2010,tad-degron-2003}.    The opposite conclusion to "competition aids oscillations" is drawn in numerical analysis of models of synthetic oscillators \cite{munteanu2010}.  

Our model displays the role of copy numbers of alleles in increasing the amplitude of oscillations (via increasing $c$).   Consequences of gene duplication have been considered in the context of dosage balance.  The ubiquity of dominance  has given rise to the hypothesis that dosage balance is favoured \cite{heterosis2010}.  On the other hand, in the case of the oscillatory dynamical states discussed, our assumptions favoured a heterozygote -- that under  some environmental (typically periodic) conditions there is selective advantage to oscillatory behaviour  -- leading to the increase in frequency of gene duplicates with genotypes with an even larger amplitude for oscillations.   On a large-scale analysis, it has been noted that there is an increased fraction \cite{chen-WGDosc08} of duplicated yeast genes as a result of whole genome duplication amongst those that cycle during metabolic oscillations \cite{tu-metab-osc-science05}, and \cite{trachana-embo2010} identify the extensive participation of paralogs in multiple rhythmic processes as a partitioning of the oscillatory feature amongst duplicates.  We leave for a future investigation the theoretical analysis of duplications in such regulatory model systems.

\appendix
\section{Model construction}

In this Appendix  we describe, first, the thermodynamic model of transcription rates, then the detailed kinetic model that facilitates the introduction of the stochastic model used for performing the simulations.  We also introduce the context dependent model which takes the influence of \emph{cis}-regulatory site information in the rate of transcription.

%
\subsection{Thermodynamic model of promoter occupancy}
Transcription factors bound to the enhancer recruit RNA polymerase
and the thermodynamic
formalism can easily accommodate different binding free energies of
protein-DNA and protein-protein interactions in a uniform manner.  In
this paper, we summarize the composite effects of transcription
factor-Mediator and Mediator-RNA polymerase binding by energy terms
$\epsilon_{A_i p}$ for each activator $A_i$.  Hereafter, we denote by
$P$ the transcriptional apparatus involving general transcription
factors, the Mediator complex and RNA Pol II.  In the standard way to
count different possible configurations (see \cite{Bintu2005a} for a
pedagogical introduction in the context of gene regulation) we introduce
Boltzmann factors for all possible configurations for binding of
activators $A_{1,2}$ and $P$ to calculate the partition function 
$Z_{tot}(P,A_1,A_2)$.  A subset of these configurations are poised for
transcription \emph{i.e.}, those with RNA polymerase or Pol II bound
to the promoter.  The rate of mRNA synthesis is taken to be
proportional to the probability of the Pol II bound promoter, which is
taken to be the ratio of the Boltzmann factors for the favourable
configurations with promoter-specific bound Pol II to
$Z_{tot}(P,A_1,A_2)$. 

The binding energies for non-specific (site-specific) binding are
denoted $\varepsilon^0 (\varepsilon^s)$, with appropriate subscripts
which identify the binding of activators or the polymerases to the
DNA.  For the case where $k$ polymerases, $l$ activators of type 1 and
$m$ activators of type 2 bind to the \emph{cis}-regulatory region of
the DNA their energy contribution is  
\begin{equation}
\begin{array}{rcl}
E^s(k,l,m)&:=&k(\varepsilon^s_{pd}+l\varepsilon_{A_1p}+m\varepsilon_{A_2p}+lm\varepsilon_{A_1A_2p})+\\
&&\;l\varepsilon^s_{A_1d}+m\varepsilon^s_{A_2d}+lm\varepsilon_{A_1A_2},
\end{array}
\end{equation}
where the subscripts indicate the protein-protein binding energies as
well, including $\varepsilon_{A_1A_2p}$ which captures the net energy
of recruitment of transcriptional machinery due to the combined action
of the activators.  If the proteins bind to non-cognate sites, the
binding energy is  
\begin{equation}
E^0(k,l,m):=k \varepsilon^0_{pd}+l\varepsilon^0_{A_1d}+m\varepsilon^0_{A_2d}.
\end{equation}
For $P$  the number of Pol II molecules, $A_{1,2}$ the number of
transcription activators of each type, we introduce
\begin{equation}
\zeta(P,A_1, A_2)=\left(N_{ns}\atop {P,A_1, A_2} \right) \displaystyle
e^{-\beta E^0(P,A_1,A_2)},
\end{equation}
where the right hand side includes in the exponent $\beta=1/(k_B T)$,
and the trinomial coefficient contains the number of 
binding sites in the genome $N_{ns}$.   We shall simplify the partition function for 0 or 1 molecules of type $P,
A_{1,2}$ bound to the relevant DNA sites 
\begin{equation}
Z_{tot}(P,A_1,A_2)=\displaystyle\sum_{(k,l,m)\in\{0,1\}^3}\zeta(P-k,A_1-l,A_2-m)e^{-\beta
  E^s(k,l,m)}
\end{equation}
using the (Stirling) approximation 
\begin{equation}
\binom{N_{ns}}{X,Y,Z}\approx \frac{N_{ns}^{X+Y+Z}}{X! Y! Z!}\,\mbox{ if }\, N_{ns}\gg X,Y,Z,
\end{equation}
and the following definitions (\ref{eq:deftherm1}), 
\begin{equation}
\label{eq:deftherm1}
\begin{array}{c}
 \alpha_i =  \displaystyle  \frac{A_i}{N_{ns}}e^{-\beta(\varepsilon^s_{A_i d}-\varepsilon^0_{A_i d})}, \quad
 r_i =  \displaystyle e^{-\beta \varepsilon_{A_i p}}, \quad i=1,2 \\   
 \rho = \displaystyle \frac{P}{N_{ns}}e^{-\beta(\varepsilon^s_{pd}-\varepsilon^0_{pd})},\;
r_{12} =  \displaystyle e^{-\beta \varepsilon_{A_1 A_2 p}}, \;
  \omega_{12} =  \displaystyle e^{-\beta \varepsilon_{A_1 A_2}}. 
\end{array}
\end{equation}
We obtain
\begin{equation}
\displaystyle \frac{Z_{tot}(P,A_1,A_2)}{\zeta(P, A_1, A_2)}= 1+\rho+(1+\rho r_1)\alpha_1 +(1+\rho r_2)\alpha_2 + (1+\rho r_{12}) \omega_{12} \alpha_1 \alpha_2,
\end{equation}
which enables us to compute the probability of occupancy of the
promoter by RNA Pol II:
\begin{equation}
\displaystyle\frac{1}{Z_{tot}(P,A_1,A_2)}\sum_{(l,m)\in\{0,1\}^2}\zeta(P-1,A_1-l,A_2-m)e^{-\beta
  E^s(1,l,m)}.
\end{equation}
Finally, we can express the transcription rate as proportional to
probability of promoter occupancy by $P$:
\begin{equation}
\label{eq:thermosummary}
\displaystyle \Phi = \frac{\rho\left(1+ r_1\alpha_1 + r_2 \alpha_2 + r_{12}\omega_{12} \alpha_1 \alpha_2  \right)}{\left(1+ \alpha_1 + \alpha_2 + \omega_{12} \alpha_1 \alpha_2  \right)+\rho\left(1+ r_1\alpha_1 + r_2 \alpha_2 + r_{12}\omega_{12} \alpha_1 \alpha_2  \right)}.
\end{equation}
Phenomenologically, such an expression is matched to experimental data on amplification or reduction of mRNA production as a function of transcription factor numbers, called the fold change function, $\psi$: 
\begin{equation}
\label{eq:thermfoldchange}
\psi(\alpha_1,\alpha_2)=\displaystyle \frac{\rho\left(1+ r_1\alpha_1 + r_2 \alpha_2 + r_{12}\omega_{12} \alpha_1 \alpha_2  \right)}{\left(1+ \alpha_1 + \alpha_2 + \omega_{12} \alpha_1 \alpha_2  \right)}.
\end{equation}
For the case of competitive binding of $A_1$ and $A_2$ to the same promoter site, we shall set $\omega_{12}=0$ hereafter.  

\subsection{Context-dependent recruitment}
\label{context1}

In order to introduce \emph{cis}-context dependent regulation beyond that captured by $A_{1,2}$ binding sites, we will need to introduce additional binding sites.  We introduce two additional  factors $A_3$ and $A_4$ but we assume that it only alters the transcriptional ability of $A_{1,2}$ by site-specific and protein-protein interactions.  We proceed in exactly the same lines as before to end up with 
\begin{equation}
\label{eq:thermfoldchange-ext0}
\psi(\alpha_1,\alpha_2,\alpha_3,\alpha_4)=\displaystyle \frac{\rho\left(1+ \sum_{i=1}^4 r_i\alpha_i + \sum_{k=3}^4 r_{ik}\omega_{ik} \alpha_i \alpha_k  \right)}{\left(1+ \sum_{i=1}^4 \alpha_i + \sum_{k=3}^4 \omega_{ik} \alpha_i \alpha_k  \right)},
\end{equation}
and the promoter occupancy is, as before,
\[
\Phi=\displaystyle \frac{1}{1+\psi^{-1}}.
\] 
In what follows we shall restrict ourselves to the situation where $A_3$ and $A_4$ does not initiate transcription on its own but merely acts as helpers to $A_{1,2}$.  Thus we can set $r_3=0=r_4$ and end up with a context dependent fold-change factor $\psi^{(c)}$
\begin{equation}
\label{eq:thermfoldchange-ext}
\psi^{(c)}(\alpha_1,\alpha_2)=\displaystyle \frac{\rho\left(1+ \alpha_1 (r_1 + r_{13}\omega_{13} \alpha_3 + r_{14}\omega_{14} \alpha_4)  +  \alpha_2 (r_2 +  r_{23}\omega_{23} \alpha_3 + + r_{24}\omega_{24} \alpha_4)  \right)}{\left(1+ \alpha_1 (1  +   \omega_{13} \alpha_3 +   \omega_{14} \alpha_4)  + \alpha_2 (1 + \omega_{23} \alpha_3 + \omega_{24} \alpha_4)\right)},
\end{equation}
which enables a greater flexibility in setting differential rates of recruitment based on both protein-protein interactions $\omega_{ij}$, sequence-dependent transcription factor-DNA binding ($\alpha_i$) as well as recruitment of the transcriptional machinery ($r_i$, $r_{ij}$). 


\subsection{Kinetic model and reduction}
\label{kinetic}

In order to perform stochastic simulations it is convenient to also
introduce a detailed kinetic scheme which reduces to the result
derived from thermodynamics under detailed balance of the fast
reactions to be indicated below.  As in the thermodynamic description, we shall describe the details of the two activator case and extend the formalism to include the third auxiliary transcription factor as that does not play a dynamical role in the paper, but merely motivates the parameterization.

For this section it is convenient to
introduce $a$ and $b$ as labels that refer to the two genes, and $A$
and $B$ to denote the two proteins that are expresed.  We shall also
refer to the quantities of $A$ and $B$-monomers by [A] and [B] and
their dimeric forms by [A2] and [B2] respectively.  Later on we shall
also introduce the variables $x$ and $y$ which are scaled versions of
[A] and [B].  To match the ($x_1, x_2$) variables in the paper, we
note that $x=x_1$ and $y=x_2$.  To connect to the thermodynamic
description in \ref{eq:thermosummary},  $\alpha_{1,2}$ will be taken
to be dimers [A2], [B2] and will be represented by $x^2$ and $y^2$ in
the Appendix , and to repeat, is called $x_{1,2}$ in
the main paper.  

The set of reactions that we consider are given in the table below.
\begin{table}
\[
\begin{array}{lcc}
\text{Reactions}&\text{Rates} &\text{Propensities}\\\hline
 2 A\rightleftarrows \text{A2} & (k_{dim,A}^b , k_{dim,A}^u) &
 (k_{dim,A}^b n_A^2/\Omega , k_{dim,A}^un_{\scriptscriptstyle  A2})  \\
  2 B\rightleftarrows \text{B2} & (k_{dim,B}^b , k_{dim,B}^u) &
  (k_{dim,B}^b n_{\scriptscriptstyle  B}^2/\Omega, k_{dim,B}^u) n_{\scriptscriptstyle  B2} \\
 \text{A2}+\text{DA}\rightleftarrows \text{DAA2} & (k_{da,A}^b ,
   k_{da,A}^u) & (k_{da,A}^b n_{\scriptscriptstyle  A2} n_{\scriptscriptstyle  DA}/\Omega ,
   k_{da,A}^un_{\scriptscriptstyle  DAA2}) \\
 \text{A2}+\text{DB}\rightleftarrows \text{DBA2} & (k_{db,A}^b ,
  k_{db,A}^u) & (k_{db,A}^b n_{\scriptscriptstyle  A2}n_{\scriptscriptstyle  DB}/\Omega ,
  k_{db,A}^u) n_{\scriptscriptstyle  DBA2}\\
 \text{B2}+\text{DA}\rightleftarrows \text{DAB2} & (k_{da,B}^b ,
   k_{da,B}^u) & (k_{da,B}^b n_{\scriptscriptstyle  DA}n_{\scriptscriptstyle  B2}/\Omega ,
   k_{da,B}^u n_{\scriptscriptstyle  DAB2}) \\
 \text{B2}+\text{DB}\rightleftarrows \text{DBB2} & (k_{db,B}^b ,
  k_{db,B}^u) & (k_{db,B}^b n_{\scriptscriptstyle  DB}n_{\scriptscriptstyle  B2}/\Omega,
  k_{db,B}^un_{\scriptscriptstyle  DBB2})\\
\text{DA}+\text{Pol}\rightleftarrows \text{DAPol} & (\rho_{a0}^+ ,
   \rho_{a0}^-) & (\rho_{a0}^+n_{\scriptscriptstyle  DA} n_{\scriptscriptstyle  Pol}/\Omega ,
   \rho_{a0}^- n_{\scriptscriptstyle  DAPol})\\
 \text{DB}+\text{Pol}\rightleftarrows \text{DBPol} & (\rho_{b0}^+ ,
   \rho_{b0}^-) & (\rho_{b0}^+n_{\scriptscriptstyle  DB} n_{\scriptscriptstyle  Pol}/\Omega ,
   \rho_{b0}^- n_{\scriptscriptstyle  DBPol}) \\
   \text{DAA2}+\text{Pol}\rightleftarrows \text{CAA2} & (\rho_{aA}^+ ,
   \rho_{aA}^-) & (\rho_{aA}^+n_{\scriptscriptstyle  DAA2} n_{\scriptscriptstyle  Pol}/\Omega ,
   \rho_{aA}^- n_{\scriptscriptstyle  CAA2})\\
 \text{DAB2}+\text{Pol}\rightleftarrows \text{CAB2} & (\rho_{aB}^+ ,
   \rho_{aB}^-) & (\rho_{aB}^+n_{\scriptscriptstyle  DAB2} n_{\scriptscriptstyle  Pol}/\Omega ,
   \rho_{aB}^- n_{\scriptscriptstyle  CAB2})\\
 \text{DBA2}+\text{Pol}\rightleftarrows \text{CBA2} & (\rho_{bA}^+ ,
   \rho_{bA}^-) & (\rho_{bA}^+n_{\scriptscriptstyle  DBA2} n_{\scriptscriptstyle  Pol}/\Omega ,
   \rho_{bA}^- n_{\scriptscriptstyle  CBA2}) \\
 \text{DBB2}+\text{Pol}\rightleftarrows \text{CBB2} & (\rho_{bB}^+ ,
   \rho_{bB}^-) & (\rho_{bB}^+n_{\scriptscriptstyle  DA} n_{\scriptscriptstyle  Pol}/\Omega ,
   \rho_{bB}^- n_{\scriptscriptstyle  DAPol})\\
 \text{DAPol}\to \text{DA}+m_a+\text{Pol} & \mu_{a,0}  & \mu_{a,0}\\
 \text{DBPol}\to \text{DB}+m_b+\text{Pol} & \mu_{b,0}  & \mu_{b,0}\\
  \text{CAA2}\to \text{DAA2}+m_a+\text{Pol} & \mu_{a,A} & \mu_{a,A} \\
 \text{CAB2}\to \text{DAB2}+m_a+\text{Pol} & \mu_{a,B}  & \mu_{a,B} \\
 \text{CBA2}\to \text{DBA2}+m_b+\text{Pol} & \mu_{b,A}  & \mu_{b,A} \\
 \text{CBB2}\to \text{DBB2}+m_b+\text{Pol} & \mu_{b,B}  & \mu_{b,B} \\
 \text{DAPol}\to \text{DA}+m_a+\text{Pol} & \mu_{a,0}  & \mu_{a,0} \\
 \text{DBPol}\to \text{DB}+m_b+\text{Pol} & \mu_{b,0} & \mu_{b,0} \\
 m_a\to A+m_a &\pi_a  &\pi_a \\
 m_a\to \emptyset  & \delta_{ma} & \delta_{ma} \\
 m_b\to B+m_b &\pi_b &\pi_b  \\
 m_b\to \emptyset  & \delta_{mb} &\delta_{mb}  \\
 A\to \emptyset  & \Delta_{A}  & \Delta_{A}\\
 B\to \emptyset  & \Delta_{B} & \Delta_{B} \\ \hline
 \text{B2}+\text{DAA2}\rightleftarrows \text{DAX} & (k_{da,X}^b ,
  k_{da,X}^u) = ({0,\infty})& ({0,\infty})\\
 \text{A2}+\text{DBB2}\rightleftarrows \text{DBX} &  (k_{db,X}^b ,
  k_{db,X}^u) =  ({0,\infty})& ({0,\infty})\\
\end{array}
\]
\caption{The molecular species/states represented in the reaction
  scheme are labelled by the genetic identities, $A$ and $B$.  $DA,
  DB$ stand for promoter regions upstream of genes $a$, $b$.  mRNA and
proteins are $m_a, A$, etc, while Pol is a shorthand for the set of
intermediates including Mediator and the RNA Pol II transcriptional
machinery.  Of key significance in this paper is the different
affinities of the transcription factors $A2$, etc bound to promoters
$DAA2$, etc have for this transcribing machinery.  The states $DAA2$
and $DAB2$ refer to the states of the promoter of gene $a$ bound by
$A2, B2$ respectively.  The set of reactions below the horizontal line
involving states DAX, CAX, DBX, CBX are those with promoters bound by both
transcription factors.  These states are excluded in the XOR case,
$k_{da,X}^b = 0 = k_{db,X}^b$.  We
have also set the basal rates of polymerase binding to the two
promoters the same, $\rho_0^{\pm}$.  The rightmost column gives the
propensities for the reactions used in the Gillespie simulation.  The
probabilities for the stochastic case are obtained by dividing the
rate constants by $N_A \Omega$ where $N_A$ is Avogadro's number and
$\Omega$ a volume.  Under this normalisation, a 1nM concentration
corresponds to approximately 1 molecule in \emph{E. coli} and 60
molecules in a mammalian cell nucleus.
}
\label{tab:reactions2}

\end{table}
Note that the reactions below the horizontal line in the table are
\emph{excluded}, and are presented only to indicate which reactions
occur with vanishing probability because of the steric inhibition of
the transcription factors. In order to relate 
the kinetic description under detailed balance to 
the thermodynamic description, we set
\begin{equation}
\label{eq:exclusion}
\omega_{12}=0
\end{equation}
in \ref{eq:thermosummary} to impose the mutual exclusivity of binding
to the enhancers.

In the reactions in Table \ref{tab:reactions2}, we make the assumption
that all the binding-unbinding events in the \emph{cis}-regulatory
regions are much faster compared to the slow processes of transcript
formation and translation.  Further, we assume detailed balance to
arrive at the same fractions for the states corresponding to the bound
configurations  CAA2, CAB2, CBA2, CBB2 (and CA, CB for basal transcription)
as in the thermodynamical formulation \ref{eq:thermosummary}.  

The rates of transcription of the mRNA species of A and B are:
\begin{equation}
\label{eq:transcriptioneqs1}
\begin{array}{rcl}
\displaystyle\frac{d}{dt}[m_a]&=&\mu_{a,0}[CA]+\mu_{a,A}[CAA2]+\mu_{a,B}[CAB2]-\delta_a
[m_a]\\
\displaystyle\frac{d}{dt}[m_b]&=&\mu_{mb0}[CB]+\mu_{b,A}[CBA2]+\mu_{b,B}[CBB2]-\delta_b
[m_b]
\end{array}
\end{equation}
where the promoters [DA], [DB] of genes A and B are bound by the RNA
Pol II and initiate transcript elongation at the different rates $\mu$.
We shall make the assumption in this paper that transcription
elongation takes place at a rate that is independendent of promoter
configuration and so we shall set
\[
\mu_{a,0}=\mu_{a,A}=\mu_{a,B}=\mu_a\;\mbox{and}\;\mu_{b,0}=\mu_{b,A}=\mu_{b,B}=\mu_b.
\]
In order to match up with the thermodynamic formalism we need to \emph{assume} detailed
balance for all the reactions where $\triangleright$ binds to/unbinds
from $\triangleleft$ to form $\bowtie$ in Table \ref{tab:reactions2}:
\begin{equation}
\label{eq:detailedbalance}
k^+[\triangleright][\triangleleft]=k^-[\bowtie]\Rightarrow [\bowtie] =
\displaystyle\frac{k^+}{k^-} [\triangleright][\triangleleft],
\end{equation}
which prompts us to introduce:
\begin{equation}
K_{dim,A}=\frac{k^u_{dim,A}}{k^b_{dim,A}},\;
K_{dim,B}=\frac{k^u_{dim,B}}{k^b_{dim,B}}\;
\end{equation}
for the dimer dissociation constants and
\begin{equation}
K_{da,A}=\frac{k^u_{da,A}}{k^b_{da,A}}, \;
K_{db,A}=\frac{k^u_{db,A}}{k^b_{db,A}},\;
K_{da,B}=\frac{k^u_{da,B}}{k^b_{da,B}},\;
K_{db,B}=\frac{k^u_{db,B}}{k^b_{db,B}}
\end{equation}
for protein-DNA binding, where the $K_{d\circ,\bullet}$ notation
stands for $\bullet$ binding to promoter of gene $\circ$.  

The GRF involves terms of the form $([TF]/K_{\scriptscriptstyle
  TF})\times r$ where $r$ incorporates the binding probabilities of
the transcription factors to the transcriptional machinery (such as
Mediator, RNA polymerases, etc) and $K_{\scriptscriptstyle TF}$ is the
dissociation constant for protein-DNA interaction (involves
exponentials of free energy of binding).  The TFs of interest in this
model are dimers $A_2$ of protein $A$ and from  
\[A+A  \rates{k_{dim,
    A}^b}{k_{dim, A}^u}A2\quad K_{dim}^{A}=\frac{k_{dim, A}^u}{k_{dim,
    A}^b}=\frac{[A]^2}{[A2]}\]. 
Using the detailed balance rapid equilibrium hypothesis, the dynamical
variables enter with products of the dissociation constants for
protein-DNA and dimerisation dissociation constants.  

We introduce 
\begin{equation}
\label{eq:rs}
r_{i0}:=\text{[Pol]}\frac{\rho^+_{i0}}{\rho^-_{i0}},\;
r_{iA}:=\text{[Pol]}\frac{\rho^+_{iA}}{\rho^-_{iA}}
,\; r_{iB}:=\text{[Pol]}\frac{\rho^+_{iB}}{\rho^-_{iB}}
\end{equation}
where $i$ could take the index values $a$ or $b$ and would reflect the
promoter dependence of the (basal) dissociation rates in question.  We further
define $\kappa_A^2={K_{dim}^A K_{da, A}}$ and
$\kappa_B^2={K_{dim}^B K_{db, B}}$ so that
\[
x=[A]/\kappa_A\;\; \text{and} \;\;
y=[B]/\kappa_B,
\] and the ratios $t^{-1}_{ij}=(K_{di,j}/K_{dj,j})$, with $t_{ii}=1$.  Thus,
$t^{-1}_{aB}=(K_{da,B}/K_{db,B})$, $t^{-1}_{bA}=(K_{db,A}/K_{da,A})$ measure
the relative strengths of binding of the transcription factors to the
promoter regions of the 2 genes using their autoregulatory binding as reference. 
Now we substitute for the bound promoter concentrations in
eq. (\ref{eq:transcriptioneqs1}) using the relations derived from detailed balance
to obtain
\begin{equation}
\label{eq:transcriptioneqs2}
\begin{array}{rcl}
\displaystyle\frac{d}{dt}[m_a]&=&\displaystyle -\delta_a
[m_a] + \mu_{a}[DA]\text{[Pol]}\left(\frac{\rho_{a0}^+}{\rho_{a0}^-}
+\right. \\
&& \left. \displaystyle\frac{\rho_{aA}^+}{\rho_{aA}^-}\frac{\text[A]^2}{K_{dim}^A
    K_{da,A}} + \frac{\rho_{aB}^+}{\rho_{aB}^-}\frac{K_{db,B}}{K_{da,B}}\frac{\text[B]^2}{K_{dim}^B K_{db,B}}  \right)    
\\
&&\\
&=& -\delta_a
[m_a] + \mu_a [DA]\left(r_{a0}+r_{aA}x^2+r_{aB}t_{aB} y^2 \right)\\
&&\\
\displaystyle\frac{d}{dt}[m_b]&=&-\delta_b
[m_b] + \displaystyle\mu_{b}[DB]\text{[Pol]}\left(\frac{\rho_{b0}^+}{\rho_{b0}^-}
+  \right. \\
&& \left. \displaystyle\frac{\rho_{bA}^+}{\rho_{bA}^-}\frac{K_{da,A}}{K_{db,A}}\frac{\text[A]^2}{K_{dim}^A
    K_{da,A}} + \frac{\rho_{bB}^+}{\rho_{bB}^-}\frac{\text[B]^2}{K_{dim}^B K_{db,B}}  \right)    
\\
&&\\
&=&-\delta_b [m_b] + \mu_b [DB]\left(r_{b0}+r_{bA}t_{bA} x^2 + r_{bB} y^2 \right).
\end{array}
\end{equation}
To obtain the expressions for the promoter fractions [DA] and
[DB], note that the total available promoters are either unoccupied,
or occupied by transcription factors and core transcriptional
machinery.  Thus, the total promoter availability is used to obtain
the bound fractions, in line with the thermodynamic description above.
These are obtained as follows:  
\begin{equation}
\label{eq:promotertotals}
\begin{array}{rcl}
[DA]_0&=&[DA]+[CA] + [DAA2]+[CAA2] +
[DAB2]+[CAB2]\\
&&\\
\Rightarrow \displaystyle \frac{[DA]_0}{[DA]}&=&\displaystyle 
 (1+\frac{\rho_{a0}^+}{\rho_{a0}^-}\text{[Pol]}) + 
 (1 +
  \frac{\rho_{aA}^+}{\rho_{aA}^-}\text{[Pol]})\frac{\text[A]^2}{K_{dim}^A
    K_{da,A}} \\
    &&\qquad\qquad\qquad\displaystyle  +
(1 +
  \frac{\rho_{aB}^+}{\rho_{aB}^-}\text{[Pol]})\frac{K_{db,B}}{K_{da,B}}\frac{\text[B]^2}{K_{dim}^B K_{db,B}}\\
&&\\
&=&(1+r_{a0}) + (1+r_{aA})x^2 + (1+r_{aB}) t_{aB} y^2 \equiv \Gamma_a \\
&&\\
{[DB]}_0&=&{[DB]}+[CB] + [DBA2]+[CBA2] +
[DBB2]+[CBB2]\\
&&\\
\Rightarrow \displaystyle \frac{[DB]_0}{[DB]}&=&\displaystyle 
 (1+\frac{\rho_{b0}^+}{\rho_{b0}^-}\text{[Pol]}) + 
 (1 +
  \frac{\rho_{bA}^+}{\rho_{bA}^-}\text{[Pol]})\frac{K_{da,A}}{K_{db,A}}\frac{\text[A]^2}{K_{dim}^A
    K_{da,A}}  \\
    &&\qquad\qquad\qquad\displaystyle + 
(1 +
  \frac{\rho_{bB}^+}{\rho_{bB}^-}\text{[Pol]})\frac{\text[B]^2}{K_{dim}^B K_{db,B}} \\
&&\\
&=&(1+r_{b0}) + (1+ r_{bA}) t_{bA} x^2 + (1+r_{bB})y^2 \equiv \Gamma_b
\end{array}
\end{equation}

We now write the set of equations determining the kinetics of
transcription and translation:
\begin{equation}
\begin{array}{rcl}
\displaystyle\frac{d}{dt}[m_a]&=&\displaystyle\mu_a [DA]_0
\frac{\displaystyle r_{a0} + r_{aA}x^2 + r_{aB}t_B^{-1}
  y^2}{\displaystyle (1+ r_{a0}) + (1+r_{aA})x^2 + (1+r_{aB})t_{aB} y^2}
-\delta_a
[m_a],\\
\displaystyle\frac{d}{dt}[A]&=&\displaystyle \pi_a [m_a] - \Delta_A [A],\\
\displaystyle\frac{d}{dt}[m_b]&=&\displaystyle\mu_b [DB]_0
\frac{\displaystyle r_{b0} + r_{bA}t_{bA}x^2 + r_{bB} y^2}{\displaystyle (1+ r_{b0}) + (1+r_{bA})t_{bA}x^2 + (1+r_{bB}) y^2}-\delta_b
[m_b],\\
\displaystyle\frac{d}{dt}[B]&=&\displaystyle \pi_b [m_b] - \Delta_B [B].
\end{array}
\end{equation}

We then introduce the assumption
that the fast decay times of mRNA compared to those of proteins
enables the translation machinery to effectively see a steady state
level of mRNA $[m_i]^{ss}$
\begin{equation}
[m_i]^{ss}=\frac{1}{\delta_i}\text{mRNA production rate}(i), \;
\text{for } i=a,b
\end{equation}  
which we substitute in the differential equations for proteins to
arrive at
\begin{equation}
\begin{array}{rcl}
\displaystyle\frac{d}{dt}x&=& [DA]_0\displaystyle \frac{\pi_a\mu_a}{\delta_a\kappa_A}\frac{\displaystyle r_{a0} + r_{aA}x^2 + r_{aB}t_{aB}
  y^2}{\displaystyle (1+ r_{a0}) + (1+r_{aA})x^2 + (1+r_{aB})t_{aB}
  y^2} - \Delta_A x\\&\equiv& c_a \varphi_a(x,y) - \Delta_A x,\\
&&\\
\displaystyle\frac{d}{dt}y&=& [DB]_0\displaystyle \frac{\pi_b\mu_b}{\delta_b\kappa_B}
\frac{\displaystyle r_{b0} + r_{bA}t_{bA}x^2 + r_{bB}
  y^2}{\displaystyle (1+ r_{b0}) + (1+r_{bA})t_{bA}x^2 + (1+r_{bB})
  y^2}-\Delta_B y\\&\equiv& c_b \varphi_b(x,y) - \Delta_B y,
\end{array}
\end{equation}
where $c_a = \frac{\pi_a\mu_a}{\delta_a\kappa_A}$ and $c_b = \frac{\pi_b\mu_b}{\delta_b\kappa_B}$.

To summarise the correspondence to the notation in the main text,  we have $(x_1,x_2)\leftrightarrow(x,y)$, $(t_{aA},t_{aB},t_{bA},t_{bB})\leftrightarrow(t_{11},t_{12},t_{21},t_{22})$ and $(r_{aA},r_{aB},r_{bA},r_{bB})\leftrightarrow(r_{11},r_{12},r_{21},r_{22})$, and similarly, the $_a$ and $_b$ subscripts correspond to $_1$ and $_2$ in the main text.

\subsection{Introducing \emph{cis}-regulatory context and subfunctionalizing mutations}
\label{context2}

This section deals with the case of two transcriptional activators $A_1$ and $A_2$ and two helper proteins, $C_1$ and $A_2$.  These helper proteins  facilitate discussions on \emph{cis}-regulatory context and parameterize such context-dependence.  For simplicity, we shall restrict ourselves to the situation where $C_k$ does not activate either of the duplicate genes by itself, but modifies the recruitment potential of $A_1$ and $A_2$ via protein-protein interactions and DNA binding affinity at the regulatory site, incorporating the roles of \emph{cis}- and \emph{trans}-effects in evolution.  Using the fold change function (\ref{eq:thermfoldchange-ext}) as reference, we define 
\begin{equation}
\psi^{(C)}_j(x_1,x_2)=\displaystyle\frac{r_{j0} + (r_{j1} + \displaystyle\sum_{k=1}^2 r_{j1C_k} t_{jC_k} x_{C_k}) t_{j1} x_1^2  + (r_{j2} + \displaystyle\sum_{k=1}^2 r_{j2C_k} t_{jC_k} x_{C_k} ) t_{j2} x_2^2 }{1 + (1 + \displaystyle\sum_{k=1}^2  t_{jC_k} x_{C_k}) t_{j1} x_1^2  + (1 + \displaystyle\sum_{k=1}^2  t_{jC_k} x_{C_k} ) t_{j2} x_2^2}
\end{equation}
which incorporates the protein-protein interactions between transcription factors $A_{1,2}$ and helper proteins $C_{1,2}$ that are significant for recruitment of the transcription machinery: $r_{jiC_k}$ stands for the affinity of the $A_i$-$C_k$ protein complex on the DNA regulatory region of gene $j$ to the transcription machinery (Mediator, Pol II, \emph{etc}).  $t^{-1}_{jC_k}=K_{dj,C_k}$ measures the protein-DNA dissociation constant of $C_k$ to the enhancer of $j$.  We introduce
\begin{equation}
\label{eq:contextualdef}
\begin{array}{rcl}
\tilde{t}_{jk} &\equiv& \displaystyle t_{jk}(1+\frac{C_1}{K_{dj,C_1}}+\frac{C_2}{K_{dj,C_2}}) = t_{jk}(1+t_{jC_1} x_{C_1}+t_{jC_2} x_{C_2}) \mbox{ and }\vspace{10pt} \\ \tilde{r}_{jk}&\equiv&\displaystyle \frac{r_{jk} + r_{jkC_1} t_{jC_1} x_{C_1} + r_{jkC_2} t_{jC_2} x_{C_2}}{1+t_{jC_1} x_{C_1} + t_{jC_2} x_{C_2}},
\end{array}
\end{equation}
which simplifies the expression for the fold-change to 
\begin{equation}
\psi^{(C)}_j(x_1,x_2)=\displaystyle\frac{r_{j0} + \tilde{r}_{j1}  \tilde{t}_{j1} x_1^2  + \tilde{r}_{j2} \tilde{t}_{j2} x_2^2 }{1 + \tilde{t}_{j1} x_1^2  + \tilde{t}_{j2} x_2^2}.
\end{equation}

To evaluate the nature of the regulatory activity of $x_1$ and $x_2$ in the presence of $C$ we compute the partial derivatives $\partial_{x_k} \Phi_i$ of the probability of occupancy of the promoter of $i$: 
\[
\Phi_i = \displaystyle \frac{1}{1+\psi_i^{-1}}.
\]
This can be split into two factors:
\[
\displaystyle \frac{\partial \Phi_i}{\partial x_k} = \frac{2 x_k}{((1+\psi_i)(1+\tilde{t}_{i1}x_1^2+\tilde{t}_{i2}x_2))^2}\left( (\tilde{r}_{ik}-r_{i0}) + \sum_{j=1}^2 (1-\delta_{kj})\tilde{t}_{ij} x_j^2(\tilde{r}_{ik}-\tilde{r}_{ij})  \right), 
\]
for $(i,k=1,2)$, where the second factor determines the sign of the regulatory activity.  The dependence of $\tilde{r}$ on protein-protein interactions $r_{ijC_k}$ and DNA binding strengths $t_{iC_k}$ illustrates how context-dependent changes in the nature of regulation can be achieved.  

In particular, we demonstrate how complementary loss of function mutations can yield the parameters for the exclusive switching circuit.   If all protein-protein interactions are modular, in that they occur due to contact interactions, we can set $r_{ij}=r_{\circ j}$ and $r_{ijC_k}=r_{\circ jC_k}$.  (We denote the locus independence, or the lack of an index, as $\circ$.) We assume that $r_{\circ 1C_1}=r_{\circ 2C_2}\equiv r_C$ and $r_{\circ 2C_1}=r_{\circ 1C_2}=\epsilon r_C$, for $\epsilon\ll 1$.  Mutations are assumed to leave the protein-protein interactions unaffected for the \emph{cis}-context cases.   To implement the subfunctionalization model we set $t_{1C_1}=t_{2C_2}\equiv t_C$, and $t_{1C_2}=t_{2C_1}=\epsilon\, t_C$; thus, two loss of binding site mutations for the helper proteins $C_{1,2}$ are assumed to take place. These interaction strengths are thus of the same order of magnitude as the binding to non-specific sites and can thus be absorbed into $r_{i0}$.  Further, we assume both helpers to be of the same concentration, to simplify description and analysis: $x_{C_1}\approx x_{C_2}=x_C$.

Making these substitutions into eq. (\ref{eq:contextualdef}) we get
\begin{equation}
\begin{array}{rcl}
\tilde{t}_{ij} &\approx& \displaystyle  = t_{ij}(1+ t_C x_{C} ) \vspace{10pt} \\ 
\tilde{r}_{jk}&\approx&\displaystyle \frac{r_{\circ k} + r_{\circ kC_j} t_{C} x_C}{1+ t_{C} x_C}. \\
\end{array}
\end{equation}
The strengths $r_{\circ kC_j}$ determine, for the heterozygous switches considered in the paper,  how the complementary strengths of protein-protein interactions for recruitment of polymerases are achieved by setting $r_{\circ 1C_1}>r_{\circ kC_2}$ and $r_{\circ 2C_2}>r_{\circ 2C_1}$.
%
%

\subsection{Duplicated auto-activator on target gene}

Here we consider the case where an activator gene $a$ activates itself, upregulating the production of protein $A$ and turns on a gene $z$ which expresses a protein $Z$.  As explained in the text, this is the typical motif that figures in selector genes or terminal selector genes \cite{hobert-terminalselectorPNAS08}.  We examine the consequences of duplicating $a$ so that now we have two copies $a_1$, $a_2$ which are mutually and self-activating and also inherit a common target site in $z$.

\begin{table}
\[
\begin{array}{lcc}
\text{Reactions}&\text{Rates} &\text{Propensities}\\\hline
 \text{A2}+\text{DZ}\rightleftarrows \text{DZA2} & (k_{da,A}^b ,
   k_{dz,A}^u) & (k_{dz,A}^b n_{\scriptscriptstyle  A2} n_{\scriptscriptstyle  DZ}/\Omega ,
   k_{dz,A}^un_{\scriptscriptstyle  DZA2}) \\
 \text{B2}+\text{DZ}\rightleftarrows \text{DZB2} & (k_{dz,B}^b ,
   k_{dz,B}^u) & (k_{dz,B}^b n_{\scriptscriptstyle  DZ}n_{\scriptscriptstyle  B2}/\Omega ,
   k_{dz,B}^u n_{\scriptscriptstyle  DZB2}) \\
\text{DZ}+\text{Pol}\rightleftarrows \text{DZPol} & (\rho_{z0}^+ ,
   \rho_{z0}^-) & (\rho_{z0}^+n_{\scriptscriptstyle  DZ} n_{\scriptscriptstyle  Pol}/\Omega ,
   \rho_{z0}^- n_{\scriptscriptstyle  DZPol})\\
   \text{DZA2}+\text{Pol}\rightleftarrows \text{CZA2} & (\rho_{zA}^+ ,
   \rho_{zA}^-) & (\rho_{zA}^+n_{\scriptscriptstyle  DZA2} n_{\scriptscriptstyle  Pol}/\Omega ,
   \rho_{zA}^- n_{\scriptscriptstyle  CZA2})\\
 \text{DZB2}+\text{Pol}\rightleftarrows \text{CZB2} & (\rho_{zB}^+ ,
   \rho_{zB}^-) & (\rho_{zB}^+n_{\scriptscriptstyle  DZB2} n_{\scriptscriptstyle  Pol}/\Omega ,
   \rho_{zB}^- n_{\scriptscriptstyle  CZB2})\\
 \text{DZPol}\to \text{DZ}+m_z+\text{Pol} & \mu_{z,0}  & \mu_{z,0}\\
  \text{CZA2}\to \text{DZA2}+m_z+\text{Pol} & \mu_{z,A} & \mu_{z,A} \\
 \text{CZB2}\to \text{DZB2}+m_z+\text{Pol} & \mu_{z,B}  & \mu_{z,B}\\
 \text{DZPol}\to \text{DZ}+m_z+\text{Pol} & \mu_{z,0}  & \mu_{z,0} \\
 m_z\to Z+m_z &\pi_z  & \pi_z \\
 m_z\to \emptyset  & \delta_{mz} & \delta_{mz} \\
 Z\to \emptyset  & \Delta_{Z}  & \Delta_{Z}
\end{array}
\label{tab:reactions3}
\]
\caption{The reactions to be added to the Table 1 to incorporate the action of the transcription factors $A2$, $B2$ on the target gene $z$.  The notation is analogous to Table 1 as well. }
\end{table}

If we impose a similar detailed balance condition to extract the kinetic equations from the above reaction scheme, we end up with the following scheme:
\begin{equation}
\frac{dx_i}{dt}=\displaystyle c_i\frac{1}{1+\psi_i^{-1}(\mathbf{x},\mathbf{r}_i,\mathbf{t}_i)}-\Delta_i x_i,
\end{equation}
where $\mathbf{x}=(x_1, x_2)$,  and $(x_1,x_2,x_3)=([A_1]/\kappa_1, [A_2]/\kappa_2, [Z])$, where $\kappa_i$ is the geometric mean of two dissociation constants $K_D$: the dissociation constant of dimerization of protein labelled by $i$ and the protein-DNA dissociation constant of protein $i$ and enhancer region of gene $i$;$\mathbf{r}_i=(r_{i0},r_{i1}, r_{i2})$ parameterizes the recruiting affinity of transcription factor $A_{1,2}$ for the transcriptional machinery at genomic locus $i$; $\mathbf{t}_i=(t_{i1},t_{i2})$ measures the relative strengths of the affinities of the transcription factors to enhancers of $i$: we set  
$t_{ij}=(K_{di,A_j}/K_{d1,A_1})$, $t_{i2}=(K_{di,A_2}/K_{d2,A_2})$, for $i=1,2,3$, so that $t_{11}=1=t_{22}$.  
The promoter occupancy probability $\varphi_i(\mathbf{x},\mathbf{r}_i,\mathbf{t}_i)$ is defined in terms of the fold change function $\psi_i(\mathbf{x},\mathbf{r}_i,\mathbf{t}_i)$ thus:
\begin{equation}
\begin{array}{rcl}
\varphi_i(\mathbf{x},\mathbf{r}_i,\mathbf{t}_i)&=&\displaystyle \frac{1}{1+\psi_i^{-1}(\mathbf{x},\mathbf{r}_i,\mathbf{t}_i)}\\
\mbox{where }\psi_i(\mathbf{x},\mathbf{r}_i,\mathbf{t}_i)&=&\displaystyle \frac{r_{i0}+r_{i1}t_{i1}x_1^2+r_{i2}t_{i2}x_2^2}{1+t_{i1}x_1^2+t_{i2}x_2^2}
\end{array}
\end{equation}
Note, for simplicity we have used $r$ and $t$ instead of $\tilde{r}$, $\tilde{t}$ that we introduced to explicitly demonstrate the source of context dependence in the previous subsection.

\section{Parameter values for numerical experiments}
\label{parameters}

The following parameters were used in the computations.  Units of
concentration in nM and time in hr.  $r_{a0}=r_{b0}=1/1000$, $\delta _a=1=\delta _b$,  $\Delta _B=1/10$,  
$\Delta _A = \Delta \times \Delta_B$, $\mu _b =  2$,  $\mu _a = \mu_b \sqrt{c}$,   $\pi _b = 30$, 
$\pi _a =  \pi_b \sqrt{c}$, $\kappa _A = \kappa_B = 5\sqrt{10}$ $K_{{da},A} = 1/2 =  K_{{db},B}$, $\text{DA}_0 = 1 = \text{DB}_0$.  $r_{ij}$ is taken to be $10\times r_0\times r$ where the factor $1\leq r\leq 100$ is the ratio $r_{ij}/r_{ik}$ which is taken as a variable, as are $c$ ($1\leq c\leq 10$) and $\Delta$ ($1\leq\Delta\leq25$).   For the switches, both heterozygous and homozygous, the ratio $\Delta$ of degradation rates is taken to be $1$ and so is $c$ for the pre-duplication genotype.

\section{Hopf bifurcation surface via elimination}
\label{elimination}

The fixed points of eq. (\ref{eq:ODEmodel}), $(x_1^*,x_2^*)=(c_1\Delta_2/c_2\Delta_1)=(c/\Delta)$ which are solutions to $c_i\varphi-\Delta_i x_i=0$.  Thus $c_2\varphi(x_2,(c/\Delta)x_2)-\Delta_2x_2=0$ is a cubic represented as $f_1$ below.  The eigenvalues of the Jacobian of the $(x_1,x_2)$ system are evaluated at $(x_1^*,x_2^*)$ to analyse stability; while negative eigenvalues imply local stability, as above, analysis
of instability for oscillatory solutions involves looking at complex eigenvalues 
$\lambda=a\pm i |b|$ of the Jacobian of the dynamical system at its fixed points.   
At the fixed point, when the real part $a$ goes from being negative (stable) to 
positive (unstable) as a  continuous function of a parameter in the system, while $|b|>0$ the system is set to
undergo a Hopf bifurcation and begins oscillating with a period
$2\pi/|b|$. For a 2-dimensional system, the sum of the eigenvalues at
the crossover value $a=0$ vanishes, making the trace of the
Jacobian matrix 0, while the determinant remains positive in our case.
The equations for the tracelessness of the Jacobian at the point $x_1=(c/\Delta)x_2$: 
\[
J_{ij}=\frac{1}{(1+\Psi)^2}(c_i\frac{\partial\Psi}{\partial x_j}-\Delta_i\delta_{ij})
\]
where the partial derivatives are listed in eq. (\ref{eq:dualreg}, reproduced below:
\begin{equation}
\begin{array}{rcl}
  \left(
\begin{array}{c}
\displaystyle
\partial/\partial \alpha_1\\
\displaystyle
\partial/\partial \alpha_2
\end{array}
\right)
 \Psi(\vect{\alpha}, \vect{t}=1,\vect{r})&=&\displaystyle\frac{(r_{1}-r_{2})}{(1+\alpha_1+\alpha_2)^2}\left(\begin{array}{c}\displaystyle
 \frac{r_{1}-r_{0}}{(r_{1}-r_{2})}+\alpha_2\\\displaystyle
 \frac{r_{2}-r_{0}}{(r_{1}-r_{2})}-\alpha_1
 \end{array}
 \right).
\end{array}
\end{equation}

Thus the condition for a Hopf bifurcation reduces to solutions of 
\[
c_1\frac{\partial\Psi}{\partial x_1}+c_2\frac{\partial\Psi}{\partial x_2}=\Delta_1+\Delta_2.
\]
Using the simplification $t_{ij}=1$, and upon substituting $x_1=(c/\Delta)x_2$, this turns out to be a quartic equation $f_2$ in $x_2$  with coefficients being complicated, but polynomial combinations of the parameters.
We compute the resultant of $f_1,f_2$ to find a non-trivial greatest common divisor (gcd) and common root of these polynomials.  The resultant is computed via the determinant of the Sylvester matrix of the polynomials $f_{1},f_{2}$:
\begin{equation}
\label{eq:ss-hopf}
\begin{array}{rcl}
f_1&=&\displaystyle\sum_{i=0}^3 s_ix_2^i \quad\mbox{ and  }\; f_2\;=\;\sum_{i=0}^4 h_ix_2^i\\
s_0&=&\Delta ^2 c_2 r_0,\\
s_1&=&-\Delta ^2 \left(1+r_0\right) \Delta _2,\\
s_2&=&\left(c^2 r+\Delta ^2\right) c_2 r_2,\\
s_3&=&-\left(c^2+\Delta ^2+c^2 r r_2+\Delta ^2 r_2\right) \Delta _2,\\
h_0&=&\Delta ^4 (1+\Delta ) \left(1+r_0\right){}^2 \Delta _2,\\
h_1&=&2 \Delta ^3 c_2 \left(r_0+c^2 r_0-r_2-c^2 r r_2\right),\\
h_2&=&-2 s_3\Delta ^2 (1+\Delta ) \left(1+r_0\right),\\
h_3&=&-2 c^2 (r-1) \Delta (\Delta^2 -1) c_2 r_2,\\
h_4&=&-s_3(1+\Delta ) 
\end{array}
\end{equation}
given by
\[
 \left(
\begin{array}{ccccccc}
 s_3 & 0 & 0 & 0 & h_4 & 0 & 0 \\
 s_2 & s_3 & 0 & 0 & h_3 & h_4 & 0 \\
 s_1 & s_2 & s_3 & 0 & h_2 & h_3 & h_4 \\
 s_0 & s_1 & s_2 & s_3 & h_1 & h_2 & h_3 \\
 0 & s_0 & s_1 & s_2 & h_0 & h_1 & h_2 \\
 0 & 0 & s_0 & s_1 & 0 & h_0 & h_1 \\
 0 & 0 & 0 & s_0 & 0 & 0 & h_0 \\
\end{array}
\right).
\]
 This resultant factors into two pieces that change sign, and hence contains a zero.  The first is
\[
\left(c^2( r_1- r_0)+\Delta ^2 (r_2-r_0)\right)^2 
\]
which vanishes for 
\[
\frac{c}{\Delta}=\sqrt{\displaystyle-\frac{r_2-r_0}{r_1-r_0}}.
\]
This condition implies that one of the two genes recruits Pol II more efficiently than the basal rate while the other's activation rate is less than that of basal transcription -- \emph{i.e.}, one is an activator, the other a repressor.  

The second factor is a complicated function $h(r, \Delta, c)$ and we omit it here.  However, we can plot the surface $h(r,\Delta, c)=0$, as shown in Figure \ref{fig:hopfsurface}.  This is the case that we examine in depth in the paper.

%
%
%
%

\bibliographystyle{vancouver}

\end{document}